\documentclass[acmsmall]{acmart}
\startPage{1}

\setcopyright{none}

\usepackage{longtable,booktabs}
\usepackage{subcaption}
\usepackage{cancel}

\usepackage{amsthm}
\usepackage{amssymb,amstext}
\usepackage{etoolbox}
\usepackage{thmtools}
\usepackage{thm-restate}
\usepackage{hyperref}
\usepackage{soul}
\usepackage{mathtools} 
\usepackage{bcprules} 
\usepackage{lipsum}
\usepackage{multicol}
\usepackage{graphicx,grffile}
\usepackage{xspace}
\usepackage{xfrac}
\usepackage{nicefrac} 
\usepackage{caption}
\usepackage{enumerate}
\usepackage{lstautogobble}
\usepackage{listings} 
\usepackage[textwidth=1.3cm]{todonotes}
\usepackage{commands} 
\usepackage{multirow}
\usepackage{makecell}
\usepackage{tabularx}
\usepackage{colortbl}
\usepackage{rotating}
\usepackage{xparse}
\usepackage{marginnote}
\usepackage{bm}
\usepackage{mathpartir}
\usepackage{tikz}

\IfFileExists{footnote.sty}{\usepackage{footnote}\makesavenoteenv{longtable}}{}

\begin{document}

\title[A Path To DOT]{A Path To DOT: Formalizing Fully Path-Dependent Types}         


\author{Marianna Rapoport}
\affiliation{
  \institution{University of Waterloo}            
  \country{Canada}                    
}
\email{mrapoport@uwaterloo.ca}          

\author{Ond\v rej Lhot\'ak}
\affiliation{
    \institution{University of Waterloo}            
    \country{Canada}                    
}
\email{olhotak@uwaterloo.ca}         

\begin{abstract}
The Dependent Object Types (DOT) calculus aims to formalize the Scala
programming language with
a focus on \textit{path-dependent types}~--- types such as
$x.a_1\dots a_n.T$ that depend
on the runtime value of a \textit{path} $x.a_1\dots a_n$ to an object.
Unfortunately, existing formulations of DOT can model only types
of the form $x.A$ which depend on \textit{variables} rather than
general paths.
This restriction makes it impossible to model nested module
dependencies.
Nesting small components inside larger ones
is a necessary ingredient of a modular, scalable language.
DOT's variable restriction thus
undermines its ability to fully formalize a variety of
programming-language features including Scala's module system,
family polymorphism, and covariant specialization.

This paper presents the pDOT calculus, which generalizes DOT to support types that depend on paths of arbitrary length,
as well as singleton types to track path equality.
We show that naive approaches to add paths to DOT make it inherently unsound, and present
necessary conditions for such a calculus to be sound. We discuss the key
changes necessary to adapt the techniques of the DOT soundness proofs so that they can be applied
to pDOT. Our paper comes with a Coq-mechanized type-safety proof of pDOT.
With support for paths of arbitrary length, pDOT can realize DOT's full potential for formalizing 
Scala-like calculi.
    
\end{abstract}




    \maketitle

\section{Introduction}\label{sec:intro}



Path-dependent types embody two universal principles of modular programming: abstraction and composition.
\[
    \underbrace{\text{path-dependent}}_{\text{composition}}
    \underbrace{\text{type}}_{\text{abstraction}}
\]
Abstraction allows us to leave values or types in a program unspecified to keep it generic and reusable.
For example, in Scala, we can define trees where the node type remains abstract.

\begin{lstlisting}[language=Scala,basicstyle=\scalalst]
trait Tree {
    type Node
    val root: Node
    def add(node: Node): Tree
}
\end{lstlisting}

\noindent
If an object \code{x} has type \code{Tree}, then the \textit{path-dependent type}
\code{x.Node} denotes the type of abstract nodes.

Composition is the ability to build our program out of smaller components.
For example, if we are interested in a specific kind of tree, say a
red-black tree, then we can refine the abstract \code{Node} type
to contain a \code{Color} type.

\begin{lstlisting}[language=Scala,basicstyle=\scalalst]
trait RedBlackTree extends Tree {
    type Node <: { type Color }
}
\end{lstlisting}
\noindent
This exemplifies composition in at least two ways: by having
\code{RedBlackTree} extend \code{Tree} we have \textit{inherited} its members;
and by nesting the refined definition of \code{Node} within
\code{RedBlackTree} we have used \textit{aggregation}.
If an object \code r is a \code{RedBlackTree}, then
the path-dependent type \code{r.root.Color}
allows us to traverse the composition and access the \code{Color} type member.

To fulfill their full potential with respect to composition, path-dependent types distinguish between paths that have different runtime values.
For example, if we have apple and orange trees, we want to disallow mixing
up their nodes.

\begin{lstlisting}[language=Scala,basicstyle=\scalalst]
val appleTree: Tree
val orangeTree: Tree
appleTree.add(/*\error{orangeTree.root}*/)  // expected: appleTree.Node, actual: orangeTree.Node
\end{lstlisting}
\noindent
Here, the type system considers \code{appleTree.Node} and
 \code{orangeTree.Node} to be distinct and incompatible types
 because they depend on the runtime values of different objects.

Furthermore, path-dependent types allow Scala to unify modules and objects,
so that the same language constructs can be used to specify the overall
structure of a program as well as its implementation details. The unification
of the module and term languages is witnessed by the following comparison
with the ML module system:
Scala objects correspond to ML modules,
classes to functors, and
interfaces to signatures~\cite{cake}.


The long struggle to formalize path-dependent types
recently led to machine-verified soundness proofs
for several variants of the Dependent Object Types (DOT)
calculus~\cite{wadlerfest,oopsla16,popl17}.
%
In spite of its apparent simplicity DOT is an expressive calculus
that can encode a variety of language features, and the discovery of
its soundness proof~was a breakthrough for the Scala community.
Insights from the proof have influenced the design of Scala~3
and helped uncover soundness bugs in Scala and Java~\cite{null}.


However, a crucial limitation is that the existing DOT calculi
restrict path-dependent types to depend only
on variables, not on general paths. That is,
they allow the type \code{x.Node} (path of length~1)
but not a longer type such as \code{r.root.Color}
(length~2).
We need to lift this restriction in order to
faithfully model Scala which does allow
general path-dependent types. More importantly,
this restriction must be lifted to fulfill the goal
of \textit{scalable component abstraction}~\cite{cake}, in which modules
of a program can be arbitrarily nested to form other, larger modules.

In this paper, we formalize and prove sound a generalization of
the DOT calculus~\cite{wadlerfest} with path-dependent
types of arbitrary length.
We call the new path-dependent calculus \pdot. Our Coq-verified proof is built on top of the proof
of~\citet{simpl}.

At this point, two questions naturally arise. Are fully path-dependent types really necessary?
That is, do they provide additional expressiveness, or are they just syntactic sugar over
variable-dependent types?
And if fully path-dependent types are in fact useful, what are the barriers to adding them to DOT?

\subsection*{Why Fully Path-Dependent Types Are Necessary}

The need for paths of arbitrary length is illustrated by the simplified excerpt from the implementation of the Scala~3 (``Dotty'') compiler in Figure~\ref{fig:compiler}.
Type references (\code{TypeRef}) are \code{Type}s that have an
underlying class or trait definition (\code{Symbol}),
while \code{Symbol}s in the language also have a \code{Type}.
Additionally, \code{TypeRef}s and \code{Symbol}s are nested in different packages,
\code{core.types} and \code{core.symbols}.
\begin{wide-rules}\noindent
    \begin{multicols}{2}
        \noindent
        {\hspace{2mm}\footnotesize Scala:\\}
        \begin{lstlisting}[language=Scala,basicstyle=\scalalst\footnotesize]]
        package dotty {
          package core {
            object types {
              class Type
              class TypeRef extends Type {
                val symb: /*\textbf{core.symbols.Symbol}*/
              }
            }
            object symbols {
              class Symbol {
                val tpe: /*\textbf{core.types.Type}*/
              }
            }
          }
        }
        \end{lstlisting} 
        {\hspace{1mm}\footnotesize DOT pseudocode:\\}
        \begin{tikzpicture}
        \node [opacity=0.85] (0,0) {
            \begin{lstlisting}[language=ML,basicstyle=\scalalst\footnotesize]]
            let dotty = new {
                val core = new {
                    val types = new {
                        type Type
                        type TypeRef = Type & {
                            val symb: /*\error{core.symbols}.Symbol*/
                        }
                    }
                    val symbols = new {
                        type Symbol = {
                            val tpe: /*\error{core.types}.Type*/
                        }
                    }
                }
            }
            \end{lstlisting}};
        \end{tikzpicture}
    \end{multicols}
    \caption{A simplified excerpt from the Dotty compiler in Scala. This code fragment cannot be expressed in DOT,
        as shown on the right side in DOT pseudocode}
    \label{fig:compiler}
\end{wide-rules}
It is impossible to express the above type dependencies in DOT, as shown on the right side of Figure~\ref{fig:compiler}
in DOT pseudocode
(Section~\ref{sec:dot} presents the actual DOT syntax and type system).
To replicate nested Scala modules, DOT uses objects and fields.
Unfortunately, we run into problems when typing
the \code{symb} field because the desired path-dependent type \code{core.symbols.Symbol}
has a path of length two.

We are then tempted to find a workaround.
One option is to try to reference \code{Symbol} as a path-dependent type of
length one: \code{symbols.Symbol} instead of \code{core.symbols.Symbol}.
However, this will not do because \code{symbols} is a field, and DOT
requires that field accesses happen through the enclosing object (\code{core}).
Another option is to move the definition of the \code{Symbol} type member to the place it is accessed from, to ensure that the path to the type member has length~1. \begin{lstlisting}[language=Scala,basicstyle=\scalalst]]
val types = new {
    type Type; /*\tnew{\textbf{\color{sh_keyword} type} Symbol}*/
    type TypeRef = Type & { val symb: /*\tnew{\color{sh_keyword}\this}*/.Symbol }
}
\end{lstlisting}
However, such a transformation would require flattening the nested structure of the program whenever we need to use path-dependent types.
This would limit encapsulation and our ability to organize a program according to its logical structure.
Yet another approach is to assign the \code{symbols} object to a variable
that is defined before the \code{dotty} object:
\begin{lstlisting}[language=ML,basicstyle=\scalalst]]
let symbols = /*\textbf{\color{sh_keyword} new}*/ { type Symbol = { val tpe: /*\error{dotty}*/.core.types.Type }} in
let dotty   = /*\textbf{\color{sh_keyword} new}*/ ...
\end{lstlisting}
This attempt fails as well, as the \code{symbols} object
can no longer reference the \code{dotty} package. For the above example this
means that a \code{Symbol} cannot have a \code{Type}
(see Section~\ref{sec:pathlimit} for details).

This real-world pattern with multiple nested modules and intricate dependencies 
between them
(sometimes even \textit{recursive} dependencies, as in our example),
leads to path-dependent types of length greater than one.
Because path-dependent types are used in DOT to formalize
features like parametric
and family polymorphism~\cite{family}, covariant specialization~\cite{virtual},
and wildcards, among others,
a version of DOT with just variable-dependent types
can only formalize these features in special cases.
Thus, to unleash the full expressive power of DOT
we need path-dependent types on paths of arbitrary length.

\subsection*{Why Fully Path-Dependent Types Are Hard}

The restriction to types dependent on variables rather than
paths is not merely cosmetic; it is fundamental.
A key challenge in formalizing the DOT calculi is
the {\em bad bounds} problem, discussed in Section~\ref{sec:badbounds}: the occurrence of a type member
in a program introduces new subtyping relationships, and these
subtyping relationships could undermine type safety in the
general case. To maintain type safety, the existing DOT calculi
ensure that whenever a type $x.A$ is in scope, any code in the
same scope will not execute until $x$ has been assigned some
concrete value; the value serves as evidence that type soundness
has not been subverted. 
As we show in Section~\ref{sec:pathbadbound}, if we allow a type to depend on a path,
rather than a variable, we must extend this property to paths:
we must show that whenever a scope admits a given path, that path will always
evaluate to some stable value.
The challenge of ensuring that the paths of type-selections always 
evaluate to a value is to
rule out the possibility that paths cyclically alias each other,
while at the same time keeping the calculus expressive enough
to allow recursion.
By contrast, the DOT calculus automatically avoids the problem of type selections 
on non-terminating paths (i.e. paths whose evaluation does not terminate)
because in DOT all paths are variables, and variables are considered normal form.

A second challenge of extending DOT with support for general paths
is to track \textit{path equality}.
Consider the following program:
\begin{lstlisting}[language=Scala,basicstyle=\scalalst]
val t1 = new ConcreteTree
val t2 = new ConcreteTree
val t3 = t2
\end{lstlisting}

A subclass of \code{Tree} such as \code{ConcreteTree} (not shown) refines
\code{Node} with a concrete type that implements some representation of nodes.
We want the types \code{t1.Node} and \code{t2.Node}
to be considered distinct even though \code{t1} and \code{t2} are both initialized to the
same expression. That way, we can
distinguish the nodes of different tree instances.
On the other hand, notice that the reference \code{t3} is initialized to be
an alias to the same tree instance as \code{t2}. We therefore want \code{t2.Node}
and \code{t3.Node} to be considered the same type.

How can the type system
tell the difference between \code{t1.Node} and \code{t2.Node}, so that the
former is considered distinct from \code{t3.Node}, but the latter is considered
the same? Scala uses \textit{singleton types} for this purpose.
In Scala, \code{t3} can be
typed with the singleton type \code{t2.type} which guarantees that it is an alias
for the same object as \code{t2}. The type system treats paths that are provably aliased
(as evidenced by singleton types) as interchangeable, so it considers \code{t2.Node} and
\code{t3.Node} as the same type.
We add singleton types to
pDOT for two reasons: first, we found singleton types useful for formalizing
path-dependent types, and second, enabling singleton types brings DOT closer to
Scala.



This paper contributes the following:
\begin{enumerate}[1)]
    \item The \textbf{\pdot} calculus, a generalization of DOT with \textbf{path-dependent 
        types of arbitrary length} that lifts DOT's type-selection-on-variables 
        restriction. Section~\ref{sec:mainideas} provides an intuition for
        \pdot's main ideas, and Section~\ref{sec:pdot} presents the calculus
        in detail.
    \item The first extension of DOT with \textbf{singleton types}, a 
        Scala feature that, in addition to tracking path equality,
        enables the method chaining pattern and
        hierarchical organization of components~\cite{cake}.
    \item A \textbf{Coq-mechanized type soundness proof} of \pdot that is based on
        the simple soundness proof by~\citet{simpl}. Our proof
        maintains the simple proof's modularity properties which makes
        it easy to extend \pdot with new features.
        We describe the proof in Section~\ref{sec:proof}
        and include its Coq formalization in the accompanying artifact.
    \item \textbf{Formalized examples},
        presented in Section~\ref{sec:examples},
        that illustrate the expressive
        power of \pdot: the compiler example from this section that uses
        general path-dependent types, a method chaining example that uses
        singleton types, and a covariant list implementation.
\end{enumerate}

The Coq proof of \pdot can be found under
\begin{center}
    \href{https://git.io/dotpaths}{https://git.io/dotpaths}
\end{center}


\section{DOT: Background and Limitations}\label{sec:background}

In this section, we survey the existing DOT calculus and discuss the challenges related to path-dependent types.

\subsection{The DOT Calculus}\label{sec:dot}

We begin by reviewing the syntax of the DOT calculus of~\citet{wadlerfest}, presented in Figure~\ref{fig:dotsynt}.
A term $t$ in DOT is a variable $x$, a value $v$, a function application $x\,y$, 
a field selection $x.a$,
or a let binding $\tLet x t u$.
The meanings of the
terms are standard. The syntax is in Administrative Normal Form (ANF),
which forces terms to be bound to variables by let bindings
before they are used in a function application or as the
base of a path.
Values are either lambda abstractions, which are standard, or objects.
An object $\tNew x T d$ defines a self variable $x$, which models the
Scala \code{this} construct, 
specifies a self-type $T$ for the object,
and
lists the field and type members~$d$ defined in the object, separated
by the intersection operator~$\wedge$. Both the type
$T$ and the member definitions~$d$ can recursively refer to the object itself 
through the self variable~$x$.

\begin{wide-rules}\noindent
    \begin{multicols}{3}\noindent
        \begin{flalign}
        x,\,y,\,z            \tag*{\textbf{Variable}}\\
        a,\,b,\,c            \tag*{\textbf{Term member}}\\
        A,\,B,\,C            \tag*{\textbf{Type member}}\\
        t,\,u\coloneqq\ &                     \tag*{\textbf{Term}}\\
        &x                                     \tag*{variable}\\
        &x.a                                    \tag*{field selection}\\
        &v                                      \tag*{value}\\
        &x\,y                                   \tag*{application}\\
        &\ttLet x t u                            \tag*{let binding}\\
        &\tag*{}\\
        v\coloneqq\ &                           \tag*{\textbf{Value}}\\
        &\tNew x T d                            \tag*{object}\\
        &\tLambda x T t                         \tag*{lambda}\\
        d\coloneqq\ &                            \tag*{\textbf{Definition}}\\
        &\set{a=t}                              \tag*{field definition}\\
        &\set{A=T}                              \tag*{type definition}\\
        &\tAnd d {d'}                           \tag*{aggregate definition}\\
        &\tag*{}\\
        S,\,T,\,U\coloneqq &                     \tag*{\textbf{Type}}\\
        &\top                                   \tag*{top type}\\
        &\bot                                   \tag*{bottom type}\\
        &\tFldDec a T                           \tag*{field declaration}\\
        &\tTypeDec A S T                        \tag*{type declaration}\\
        &x.A                                    \tag*{type projection}\\
        &\tAnd S T                              \tag*{intersection}\\
        &\tRec x T                              \tag*{recursive type}\\
        &\tForall x S T                         \tag*{dependent function}
        \end{flalign}
    \end{multicols}
    \caption{Abstract syntax of DOT by~\citet{wadlerfest}}  
    \label{fig:dotsynt}
\end{wide-rules}

A DOT type is one of the following:
\begin{itemize}
    \item A dependent function type $\tForall x S T$ characterizes functions
    that take an argument of type $S$ and return a result of type $T$.
    The result type $T$ may refer to the parameter $x$.
    \item A recursive type $\tRec x T$ is the type of an object $\tNew x T d$.
    The type $T$ describes the members of the object, and may refer to
    the recursive self variable $x$.
    \item A field declaration type $\tFldDec a T$ classifies objects that have
        a field named $a$ with type $T$.
    \item A type-member declaration type $\tTypeDec A S U$ classifies objects
    that declare a type member $A$ with the constraints that $A$ is a
    supertype of $S$ and a subtype of $U$.
    \item A type projection $x.A$ selects the member type $A$ from the object referenced
    by the variable~$x$.
    \item An intersection type $\tAnd S T$ is the greatest subtype of both $S$ and $T$.
        Unlike some other systems with intersection types, DOT does not define
        any distributive laws for intersections.
    \item The top ($\top$) and bottom ($\bot$) types are the supertype and subtype of all types,
    respectively.
\end{itemize}

In the following, we will write $\nu(x)d$ instead of $\tNew x T d$ if the self type of an object is not important.
If a self variable is not used in the object we will denote it with an underscore: $\nu(\_)d$.


DOT's operational semantics is presented in Figure~\ref{fig:red2} on Page~\pageref{fig:red2}.
The reduction relation operates on terms whose free variables are bound in a value environment $\sta$
that maps variables to values.
In DOT, variables and values are considered normal form, i.e. they are irreducible.
In particular, objects $\tNew x T d$ are values, and the fields of an object are not evaluated
until those fields are selected. DOT fields are thus 
similar to Scala's
\code{lazy val}s which declare immutable, lazily evaluated values (however,
\code{lazy val}s differ from DOT fields because DOT does not memoize its fields).
\footnote{
    Evaluating fields strictly would require DOT to introduce a field initialization order which would complicate
    the calculus. DOT is designed to be a small calculus that focuses on formalizing path-dependent types
    and it deliberately leaves initialization as an open question. For a DOT with constructors and strict fields,
    see~\citet{constructors}.
}

Please refer to~\citet{wadlerfest} for a full
explanation of DOT's type rules which are presented in Figure~\ref{fig:dottyping}.
\begin{wide-rules}
\textbf{Term typing}
\begin{multicols}{2}

\infrule[Var\dotpref]
  {\G(x)=T}
  {\typDft x T}
  
\infrule[\alli\dotpref]
  {\typ {\extendG x T} t U
    \andalso
    x\notin\fv T}
  {\typDft{\tLambda x T t}{\tForall x T U}}

\infrule[\alle\dotpref]
  {\typDft x {\tForall z S T}
    \andalso
    \typDft y S}
  {\typDft {x\, y} {\tSubst z y T}}

\infrule[\newi\dotpref]
  {\typ {\extendG x T} d T}
  {\typDft {\tNew x T d} {\tRec x T}}
  
\infrule[\newe\dotpref]
  {\typDft x {\tFldDec a T}}
  {\typDft {x.a} T}

\infrule[Let\dotpref]
  {\typDft t T
      \\
    \typ {\extendG x T} u U
    \andalso
    x\notin\fv U}
  {\typDft {\ttLet x t u} U}

\infrule[\reci\dotpref]
  {\typDft x T}
  {\typDft x {\tRec x T}}

\infrule[\rece\dotpref]
  {\typDft x {\tRec z T}}
  {\typDft x {\tSubst z x T}}

\infrule[\andi\dotpref]
  {\typDft x T
    \andalso
    \typDft x U}
  {\typDft x {\tAnd T U}}
\newrulefalse

\infrule[Sub\dotpref]
  {\typDft t T
    \andalso
    \subDft T U}
  {\typDft t U}

\end{multicols}

\textbf{Definition typing}
\begin{multicols}{2}
  
\infrule[Def-Trm\dotpref]
  {\typDft t U}
  {\typDft {\set{a=t}} {\tFldDec a U}}

\infax[\deftyp\dotpref]
  {\typDft {\set{A=T}} {\tTypeDec A T T}}  

\infrule[AndDef-I\dotpref]
  {\typDft {d_1} {T_1}
    \andalso
    \typDft {d_2} {T_2}
    \\
    \dom{d_1},\,\dom{d_2}\text{ disjoint}}
  {\typDft {\tAnd {d_1} {d_2}} {\tAnd {T_1} {T_2}}}

\end{multicols}

\textbf{Subtyping}

\begin{multicols}{3}
    
\infax[Top\dotpref]
  {\subDft T \top}

\infax[Bot\dotpref]
  {\subDft \bot T}

\infax[Refl\dotpref]
  {\subDft T T}
  
\infrule[\fldsub\dotpref]
  {\subDft T U}
  {\subDft {\tFldDec a T} {\tFldDec a U}}

\infrule[\suband\dotpref]
  {\subDft S T
    \andalso
    \subDft S U}
  {\subDft S {\tAnd T U}}

\infax[\andonesub\dotpref]
  {\subDft {\tAnd T U} T}

\infax[\andtwosub\dotpref]
  {\subDft {\tAnd T U} U}
  
\infrule[\subsel\dotpref]
  {\typDft x {\tTypeDec A S T}}
  {\subDft S {x.A}}

\infrule[\selsub\dotpref]
  {\typDft x {\tTypeDec A S T}}
  {\subDft {x.A} T}

\infrule[Trans\dotpref]
  {\subDft S T
    \andalso
    \subDft T U}
  {\subDft S U}

\end{multicols}
\begin{multicols}{2}

\infrule[\typsub\dotpref]
  {\subDft {S_2} {S_1}
    \\
    \subDft {T_1} {T_2}}
  {\subDft {\tTypeDec A {S_1} {T_1}} {\tTypeDec A {S_2} {T_2}}}

\infrule[\allsub\dotpref]
  {\subDft {S_2} {S_1}
    \\
    \sub {\extendG x {S_2}} {T_1} {T_2}}
  {\subDft {\tForall x {S_1} {T_1}} {\tForall x {S_2} {T_2}}}
\end{multicols}

\caption{DOT Type Rules \citep{wadlerfest}}
  \label{fig:dottyping}

\end{wide-rules}

\subsubsection{Recursion Elimination}\label{sec:recelim}


The recursion-elimination rule is of particular interest
in the context of paths:
\noindent
{\eqfontsize
	\infrule[Rec-E$_{\text{DOT}}$]
	{\typDft x {\tRec z T}}
	{\typDft x {\tSubst z x T}}}

\noindent An object in DOT can recursively refer back to itself through the
self variable.
For example,
the $a$ member of the following object evaluates to the object itself
(see the Scala version on the right):
\begin{table}[h]
\begin{tabularx}{\textwidth}{XX}
    {\small $\ttLet x {\nu(\hleq z) \set{a=\hleq z}} \dots$}
        &\lstinline[language=Scala,basicstyle=\scalalst]!val x = new \{ val a = this \}!
    \end{tabularx}
\end{table}

\noindent Since types in DOT and Scala are dependent, the type of an object
can also refer to the object itself:
\begin{table}[h]
	\begin{tabularx}{\textwidth}{XX}
		{\small $\Let\ y =\nu(\hleq z) \set{A=T}\wedge$}
		&\lstinline[language=Scala,basicstyle=\scalalst]!val y = new \{ type A = T!\\
		{\small$\phantom{\Let\ y =\nu(\hleq z)}\set{a=\tLambda x {\hleq{z.A}} x}\ \In\ \dots$}
		&\lstinline[language=Scala,basicstyle=\scalalst]!\ \ \ \ \ \ \ \ \ \ \ \ \ \ val a = (x: this.A) => x \}!
	\end{tabularx}
\end{table}
    
\noindent
The type of the field $a$ depends on the type of the object
containing it.
In DOT, this is expressed using a recursive type. The type
of our example object is
$\tRec {\hleq z} {\tAnd{\tTypeDec A T T}{\tFldDec a {\tForall x {\hleq{z.A}} {\hleq{z.A}}}}}$.

Given the let binding above, what should be the type of~$y.a$?
In the recursive type of $y$,
the~type of the field $a$ is $\tForall x {z.A} {z.A}$.
However, because the self variable $z$ is in scope only
inside the object itself, the type $\tForall x {z.A} {z.A}$ 
does not make sense
outside the object and cannot be used to type~$y.a$.
In the field selection, however, we have a name
other than $z$ for the object itself, the name~$y$. Therefore,
we can open the recursive type by replacing the
self variable~$z$ with the external name of the object~$y$,
giving $y$ the type
$\tAnd{\tTypeDec A {\tSubst z y T} {\tSubst z y T}}{\tFldDec a {\tForall x {\hleq{y.A}} {\hleq{y.A}}}}$.
This is achieved by the recursion elimination typing rule.
Now the path~$y.a$ can have the type
$\tForall x {y.A} {y.A}$.
Notice that recursion elimination is possible only when
we have a variable such as $y$ as an external name~for an object.

Just as we need to apply recursion elimination to the type of $y$
before we can type a field selection $y.a$, we must also do the
same before we can use a type-member selection $y.A$ (specifically,
to conclude that $\tSubst z y T <: y.A <: \tSubst z y T$). The recursion elimination
is necessary because the type $T$ could also refer to the self
variable $z$, and thus may not make any sense outside of the
object. Recursion elimination replaces occurrences of $z$ in the
type $T$ with the external name $y$, so that the resulting type
is valid even outside the object.
When we add path-dependent types to the calculus, an important
consideration will be recursion elimination on paths rather than just variables.

\subsection{Path Limitations: A Minimal Example}\label{sec:pathlimit}

Consider the following example DOT object in which
a type member $B$ refers to a type member $A$
that is nested inside the definition of a field $c$:
\begin{multicols}{2}\noindent
\openup-2\jot
\begin{align*}
    \Let\ x=\nu&(z)\\
               &\set{c=\nu(\_)\set{A=z.B}}\wedge\\
               &\set{B=\error{z.c.A}} \ \In\dots
\end{align*}
\begin{lstlisting}[language=Scala,basicstyle=\scalalst]
val x = new { z /*$\Rightarrow$*/
    val c: { type A } = new { type A = z.B }
    type B = z.c.A }
\end{lstlisting}
\end{multicols}
In the example, to reference the field $c$, we must first select the field's enclosing object~$x$
through its self variable~$z$. As a result, the path to $A$ leads through $z.c$
which is a path of length two.
Since DOT does not allow paths of length two, this definition of $B$ cannot be expressed in DOT without flattening the program structure so that all fields and type members
become global members of one top-level object.

In the introduction, we illustrated how one might attempt to express the above in DOT
by decomposing the path of length two into dereferences
of simple variables, which would either lead to invalid programs or
require flattening the program structure.
We could try other ways of let-binding the inner objects to variables before defining
the enclosing object, but all such attempts are doomed to failure (unless
we are willing to give up object nesting). A sequence
of let bindings imposes a total ordering on the objects and restricts an object to refer
only to objects that are defined before it. In the presence of recursive references between
the objects, as in this example, no valid ordering of the let bindings is possible
while maintaining a nested object structure.
To avoid this we could also try to transform the local variables into recursively
defined fields of
another object $z'$,
since the order in which fields are declared does not matter. However, then $A$ would again need to refer to $B$ through
$z'.x.B$ (or $z'.y.B$) which has a path of length 2.

%
%
%
\subsection{Challenges of Adding Paths to DOT}\label{sec:challenges}
If restricting path-dependent types exclusively to variables limits the expressivity of DOT then why does the
calculus impose such a constraint?
Before we explain the soundness issue that makes it difficult to extend DOT with paths
we must first review the key challenge that makes
it difficult to ensure soundness of the DOT calculus.

\subsubsection{Bad Bounds}\label{sec:badbounds}
Scala's abstract type members make it possible to define custom
subtyping relationships between types. This is a powerful but tricky
feature. For example, given any types $S$ and $U$, consider the function
$\tLambda x {\tTypeDec A S U} t$. In the body of the function, we can
use $x.A$ as a placeholder for some type that is a supertype of $S$ and
a subtype of $U$. Some concrete type will be bound to $x.A$ when the
function is eventually called with some specific argument.
Due to transitivity of subtyping, the constraints on $x.A$ additionally
introduce an assumption inside the function body that $S <: U$,
because $S <: x.A <: U$ according to the type rules \rn{\subsel$_{\text{DOT}}$} and \rn{\selsub$_{\text{DOT}}$}:
\vspace{1mm}
\begin{multicols}{2}\noindent
    \eqfontsize
    \infrule[\subsel\dotpref]
    {\typDft x {\tTypeDec A S T}}
    {\subDft S {x.A}}
    \infrule[\selsub\dotpref]
    {\typDft x {\tTypeDec A S T}}
    {\subDft {x.A} T}
\end{multicols}
\vspace{1mm}
However, recall that $S$ and $U$ are 
arbitrary types, possibly with no existing subtyping
relationship. The key to soundness is that although the function
body is type-checked under the possibly unsound assumption $S <: U$,
the body executes only when the function is called, and calling
the function requires an argument that specifies a concrete
type $T$ to be bound to $x.A$. This argument type must satisfy
the constraints $S <: T <: U$. Thus, the argument type
embodies a form of \emph{evidence}
that the assumption $S <: U$ used to type-check the function
body is actually valid.

More generally, given a term $t$ of type ${\tTypeDec A S U}$,
we can rule out the possibility of bad bounds caused by the use
of a dependent type $t.A$
if there exists some object with the same type ${\tTypeDec A S U}$.
This is because the object must bind the type member $A$ to
some concrete type $T$ respecting the subtyping constraints $S <: T$
and $T <: U$, so the object is evidence that $S <: U$.

Existing DOT calculi ensure that whenever some variable~$x$ of type~$T$ is
in scope in some term~$t$, the term reduces only after $x$ has already
been assigned a value. The value assigned to~$x$ is evidence that
$T$ does not have bad bounds.
To ensure that any code that uses the type $x.A$ executes only after
$x$ has been bound to a value of a compatible type, DOT employs a strict
operational semantics.
A variable $x$ can be introduced by one of the three binding constructs:
$\tLet x t u$, $\tLambda x T t$, or $\tNew x T d$. In the first case,
$x$ is in scope within $u$, and the reduction semantics requires that before
$u$ can execute, $t$ must first reduce to a value with the same type as $x$.
In the second case, $x$ is in scope within $t$, which cannot execute until
an argument value is provided for the parameter $x$. In the third case,
the object itself is bound to the self variable~$x$.
In summary, the semantics ensures that by the time that evaluation reaches
a context with $x$ in scope, $x$ is bound to a value, and therefore $x$'s
type does not introduce bad bounds.

The issue
of bad bounds has been discussed thoroughly in many of the
previous papers about DOT~\cite{fool12,oopsla14,wadlerfest,simpl}.

%

\subsubsection{Naive Path Extension Leads to Bad Bounds}\label{sec:pathbadbound}
When we extend the type system with types $p.A$ that depend on paths
rather than variables, we must take similar precautions to control
bad bounds. 
If a path $p$ has type ${\tTypeDec A S U}$ and
some normal form $n$ also has this type,
then $n$ must be an object
that binds to type member $A$ a type $T$ such that $S <: T <: U$.

However, not all syntactic paths in DOT have this property.
For example, in an object $\nu(x)\set{a=t}$,
where $t$ can be an arbitrary term, $t$ could loop instead
of reducing to a normal form of the same type.
In that case, there is no guarantee that a value of the type
exists, and it would be unsound to allow the path $x.a$ as
the prefix of a path-dependent type $x.a.A$.

The following example, in which a function $x.b$ is typed as an object (a record with field $c$),
demonstrates this unsoundness (the Scala version cannot be typechecked):

\begin{multicols}{2}\noindent
	\openup-2\jot
	{\footnotesize\begin{align*}
		\nu(x: &\\
		&\tFldDec a {\tTypeDec C {(\tForall y \top \top)} {\tFldDec c \top}} &&\hspace{-4mm}\wedge\tFldDec b {\tFldDec c \top} )\\
		&\set{a=x.a}&&\hspace{-4mm}\wedge\set{b=\tLambda y \top y}
		\end{align*}}
	\begin{lstlisting}[language=Scala,basicstyle=\scalalst\footnotesize]
	new {
	  lazy val a: {type C >:Any/*$\Rightarrow$*/Any <:{val c: Any}} = a
	  lazy val b: {val c: Any} = (y: Any) /*$\Rightarrow$*/ y
	}
	\end{lstlisting}
\end{multicols}
\noindent
Here, $x.b$ refers to a function $\tLambda{y}{\top}{y}$ of type $\tForall y \top \top $.
If we allowed such a definition, the following would hold:
$
    \tForall y \top \top <:x.a.C<:\tFldDec c \top
$.
Then by subsumption, $x.b$, a function, has type $\tFldDec c \top$ and therefore it must be an object.
To avoid this unsoundness, we have to rule out the type selection $x.a.C$
on the non-terminating path $x.a$.

In general, if a path $p$ has a field declaration type $\tFldDec a T$,
then the extended path $p.a$ has type~$T$, but we do not
know whether there exists a value of type $T$ because $p.a$ has not yet reduced
to a variable. Therefore, the
type $T$ could have bad bounds, and we should not allow the path
$p.a$ to be used in a path-dependent type $p.a.A$.


The main difficulty we encountered in designing \pdot was to ensure that
type selections occur only on
terminating paths while ensuring that the calculus still permits
non-terminating paths in general, since that is necessary to
express recursive functions and maintain Turing completeness of the calculus.
%

\section{Main Ideas}\label{sec:mainideas}
In this Section, we outline the main ideas that have shaped our definition of pDOT.
The pDOT calculus that implements these ideas in full detail will be presented in Section~\ref{sec:pdot}.

\subsection{Paths Instead of Variables}\label{sec:pathsvar}

To support fully path-dependent types, our calculus needs to support paths in all places where DOT permitted variables.
Consider the following example:

\begin{table}[h]
    \begin{tabularx}{\textwidth}{XX}
        {\small $\ttLet {x} {\nu(y)\set{a=\nu(z)\set{B=U}}} {x.a}$}
        &\lstinline[language=Scala,basicstyle=\scalalst]!val x = new \{ val a = new \{ type B = U \}\}; x.a!
    \end{tabularx}
\end{table}
In order to make use of the fact that $U<:x.a.B<:U$, we need a type rule that reasons about path-dependent types. In DOT,
this is done through the  \rn{\selsub\dotpref} and \rn{\subsel\dotpref} rules mentioned in Section~\ref{sec:badbounds}.
Since we need to select $B$ on a path $x.a$ and not just on a variable $x$, we need to extend the rules (merged into one
here for brevity) to support paths:
\vspace{1mm}
\begin{multicols}{3}
    \openup-2\jot
    {\eqfontsize
        \infrule[$<:$-\selsub\dotpref]
        {\typDft x {\tTypeDec A S T}}
        {\subDft S {x.A<:T}}}
    \begin{center}{\Large $\Rightarrow$}\end{center}
    {\eqfontsize \infrule[$<:$-\selsub]
        {\typDft {\new p} {\tTypeDec A S T}}
        {\subDft S {{\new p}.A<:T}}}
\end{multicols}
\vspace{1mm}
However, before we can use this rule we need to also generalize the recursion elimination rule \rn{Rec-E\dotpref}.
In the above example, how do we obtain the typing $\typDft {x.a} {\tTypeDec B U U}$?
The only identifier of the inner object is $x.a$, a path.
The type of the path is 
$\tRec {z}{\tTypeDec B U U}$.
In order to use the type member $B$, it is necessary
to specialize this recursive type, replacing the recursive
self variable $z$ with the path $x.a$. This is necessary
because the type $U$ might refer to the self variable $z$,
which is not in scope outside the recursive type.
Thus, in order to support path-dependent types, it is necessary
to allow recursion elimination on objects identified by
paths:
\vspace{1mm}
\begin{multicols}{3}
    \openup-2\jot
    {\eqfontsize
        \infrule[Rec-E\dotpref]
        {\typDft x {\tRec y T}}
        {\typDft x {\tSubst y x T}}}
    \begin{center}{\Large $\Rightarrow$}\end{center}
    {\eqfontsize \infrule[Rec-E]
        {\typDft {\new p} {\tRec y T}}
        {\typDft {\new p} {\tSubst y {\new p} T}}}
\end{multicols}
\vspace{1mm}
By similar reasoning, we need to generalize all DOT variable-typing rules to path-typing rules.

\subsection{Paths as Identifiers}\label{sec:pathsid}

A key design decision of \pdot is to let \textit{paths represent object identity}.
In DOT, object identity is represented by variables, which works out because variables are irreducible.
In \pdot, \emph{paths} are
\textit{irreducible},
because reducing paths would strip objects of their identity and break preservation.

\subsubsection{Variables are Identifiers in DOT}
In the DOT calculus by~\citet{wadlerfest}, variables do not reduce to values for two reasons:
\begin{itemize}
    \item \textit{type safety}: making variables irreducible is necessary to maintain preservation, and
    \item \textit{object identity}: to access the members of objects (which can recursively reference 
        the object itself), objects need to have a name;
        reducing variables would strip objects of their identity.
\end{itemize}
\newcommand{\btype}{\tFldDec a {\tRec z {\tTypeDec{B}{\tSubst y x U}{\tSubst y x U}}}}
\newcommand{\btypebad}{\tFldDec a {\tRec z {\tTypeDec{B}{\tSubst y {\error x} U}{\tSubst y {\error x} U}}}}
If variables in DOT reduced
to values, then in the previous example program, $x$ would reduce to $v=\nu(y)\set{a=\nu(z)\set{B=U}}$.
To maintain type preservation, for any type $T$ such that $\typDft x T$,
we also must be able to derive $\typDft v T$.
Since $\typDft x {\tRec y {\tFldDec a {\tRec z {\tTypeDec{B}{U}{U}}}}}$, by recursion elimination \rn{Rec-E\dotpref},
$\typDft x \tFldDec a {\tRec z {\tTypeDec{B}{\hleq{\tSubst y x U}}{\hleq{\tSubst y x U}}}}$.
Does $v$ also have that type? No!

{\eqfontsize
    \begin{align*} 
    \hspace{-20mm}
    \inferrule*[Right=\parbox{2cm}{{\scriptsize \rn{preservation\dotpref}}}]
    {\inferrule*[Right=\parbox{2cm}{{\scriptsize \rn{Rec-E\dotpref}}}]
        {\typDft {x} {\tRec y {\tFldDec a {\tRec z {\tTypeDec{B}{U}{U}}}}}}
        {\typDft {x} {\btype}}
    \qquad\qquad\quad
    \sta\colon\G\qquad\quad
    \inferrule*[Right=\parbox{2cm}{{\scriptsize \rn{{\tiny\color{red} Hypothetical Var\dotpref}}}}]
        {\sta(x)=v}
        {{\color{red} \reductionDft {x} {v}}}
    }
    {\typDft {v} {\btypebad}}
    \end{align*}}
The value $v$ has only the \textit{recursive} type
$\tRec y {\tFldDec a {\tRec z {\tTypeDec{B}{U}{U}}}}$.
Since $v$
is no longer connected to any specific name,
no recursion elimination is possible on its type.
In particular, it does not make sense to give this
value the type $\tFldDec a {\tRec z {\tTypeDec{B}{\tSubst y x U}{\tSubst y x U}}}$ because this
type refers to $x$, but after the reduction, the value is no longer associated
with this name.

The example illustrates that in DOT, variables represent the identity of objects.
This is necessary in order to access an object's members: object members can reference the object itself,
for which the object needs to have a name.

\subsubsection{Paths are Identifiers in \pdot}\label{sec:pathsnormalform}

In \pdot, \textit{paths} represent the identity of objects and therefore they must be irreducible.
Similarly to DOT, reducing paths would lead to unsoundness and strip nested objects of their identity.
Making paths irreducible means that in \pdot, we cannot have an analog of DOT's field selection rule~\rn{Proj\dotpref}.
%

\newcommand{\Btype}{\tTypeDec B {\tSubst z {x.a} U} {\tSubst z {x.a} U}}
\newcommand{\Btypebad}{\tTypeDec B {\tSubst z {{\error{x.a}}} U} {\tSubst z {{\error{x.a}}} U}}

Consider the field selection
$x.a$ from the previous example. What is its type?
By recursion elimination, $x.a$ has the
type $\Btype$.
If \pdot had a path-reduction rule \rn{Proj} analogous to DOT's \rn{\redDotProj}, then $x.a$ would reduce
to $\nu(z) {\set{B = U}}$. However, that value does not have the type $\Btype$; it only has the recursive type
$\tRec {z}{\tTypeDec B U U}$.

{\eqfontsize
    \begin{align*} 
        \hspace{-20mm}\inferrule*[Right=\parbox{2cm}{{\scriptsize \rn{preservation}}}]
            {\inferrule*[Right=\parbox{2cm}{{\scriptsize \rn{Rec-E}}}]
                {\typDft {x.a} {\tRec z {\tTypeDec B U U}}}
                {\typDft {x.a} {\Btype}}
             \qquad\qquad\quad
             \sta\colon\G\qquad\quad
             \inferrule*[Right=\parbox{2cm}{{\scriptsize \rn{{\tiny\color{red} Hypothetical Proj}}}}]
                {\sta(x)=\nu(y)\set{a=\nu(z)\set{B=U}}}
                {{\color{red} \reductionDft {x.a} {\nu(z)\set{B=U}}}}
            }
            {\typDft {\nu(z)\set{B=U}} {\Btypebad}}
    \end{align*}}
The reduction step from
$x.a$
to
$\nu(z) {\set{B = U}}$
caused the object to lose its name. Since the non-recursive
type of the term depends on the name, the loss of the name
also caused the term to lose its non-recursive type.
This reduction step violates type preservation and type soundness.

%

\subsubsection{Well-Typed Paths Don't Go Wrong}\label{sec:welltypedpaths}
If \pdot programs can return paths without reducing them to values,
could these paths be nonsensical?
The type system ensures that they cannot. In particular, we ensure that
if a path~$p$ has a type then $p$ either identifies some value,
and looking up $p$ in the runtime configuration terminates,
or $p$ is a path that cyclically aliases other paths (see below).
Additionally, as we will see in Section~\ref{sec:lookupterminates}, the
\pdot safety proof ensures that if a path has a function or object type,
then it can be looked up to a value;
if $p$ can \textit{only} be typed with a singleton type (or $\top$),
then the lookup will loop.
%

When we make paths a normal form, we also have to generalize DOT's ANF syntax
to use paths wherever DOT uses variables. For example, function application has the form $p\,q$
where $p$ and $q$ are paths rather than $x\,y$ where $x$ and $y$ are variables. As a result,
all the DOT reduction and typing rules that operate on variables are generalized to paths in \pdot.

\subsection{Path Replacement}\label{sec:replacement}

We introduce a \textit{path replacement} operation for types
that contain paths which reference the same object. 
If a path $q$ is assigned to a path $p$ then $q$ \textit{aliases} $p$.
In the tree example from Section~\ref{sec:intro},
\code{t3} aliases \code{t2}, but
\code{t1} does not alias \code{t2},
even though they identify syntactically equal objects.

If $q$ is an alias of $p$ we want to ensure that we can use $q$ in the same way as $p$.
For example, any term that has type $T\to p.A$ should also have the type $T \to q.A$,
and vice versa.
In \pdot, we achieve this by introducing a subtyping relationship between \textit{equivalent types}:
if $p$ and $q$ are aliases, and a type $T$ can be obtained from type $U$ by \textit{replacing} instances of $p$ in $U$ with $q$
then $T$ and $U$ are equivalent.
For example, $T\to q.A$ can be obtained from $T\to p.A$ by replacing $p$ with $q$, and these types are therefore equivalent.
We will precisely define the replacement operation in Section~\ref{sec:typing}.

\subsection{Singleton Types}\label{sec:sngl}

To keep track of path aliases in the type system we use \textit{singleton types}.

Suppose that a \pdot program assigns the path $q$ to $p$, and that a type $T$ can be obtained from $U$ by replacing
an instance of $p$ with $q$.
How does the type system know that $T$ and $U$ are equivalent?
We could try passing information about the whole program throughout the type checker. However, that would make
reasoning about types depend on reasoning about values, which would make typechecking more complicated and less modular~\cite{simpl}.

Instead, we ensure that the type system keeps track
of path aliasing using singleton types, an existing Scala feature.
A singleton type of a path $p$, denoted $\single p$, is a type that is inhabited
\textit{only} with the value that is represented by $p$.
In the tree example from Section~\ref{sec:intro}, to tell the type system that \code{t3} aliases \code{t2}, we 
ensure that \code{t3} has the singleton type \code{t2.type}.
This information is used to allow subtyping between aliased paths, and to allow such paths to be
typed with the same types, as we will see in Section~\ref{sec:typing}.

In \pdot, singleton types are an essential feature that is necessary to encode fully path-dependent types.
However, this makes \pdot also the first DOT formalization of Scala's singleton types. In Section~\ref{sec:examples},
we show a \pdot encoding of an example that motivates this Scala feature.

\subsection{Distinguishing Fields and Methods}\label{sec:fieldsmethods}
Scala distinguishes between fields (\code{val}s, immutable fields that are
strictly evaluated at the time of object initialization)
and methods (\code{def}s, which are re-evaluated at each invocation).
By contrast, DOT unifies the two in the
concept of a term member. Since the distinction affects which paths are legal in Scala, we
must make some similar distinction in pDOT.
Consider the following Scala program:
\begin{lstlisting}[language=Scala,basicstyle=\scalalst]
val x = new {
    val a: { type A } = t/*$_a$*/
    def b: { type B } = t/*$_b$*/ }
val y: x.a.A
val z: /*\error{x.b.B}*/
\end{lstlisting}
Scala allows path-dependent types only on \textit{stable} paths~\cite{stable}.
A \code{val} can be a part of a stable path but a \code{def} cannot. Therefore,
the type selection \code{x.a.A} is allowed but \code{x.b.B} is not.

DOT 
unifies the two concepts in one:

\vspace{-2mm}
\begin{align*}
    \Let\ x = \nu(x)
      \set{a=t_a}
      \wedge\set{b=t_b}\hspace{7mm}\In\dots
\end{align*}
However, this translation differs from Scala in the order of evaluation.
Scala's fields, unlike DOT's, are fully evaluated to values when the object is constructed. Therefore, a more accurate translation of this example would be as follows:

\vspace{-2mm}
\begin{align*}
&\hleq{\Let\ a' =t_a}&&\hspace{-20mm}\In\\ 
&\Let\ x = \nu(x)
\set{a=\hleq{a'}}
\wedge\set{b=\hleqq{\lambda(\_).}\,t_b}&&\hspace{-20mm}\In\dots
\end{align*}
This translation highlights the fact that although Scala can initialize \code{x.a} to an arbitrary term,
that term will be already reduced to a value before evaluation reaches a context that contains \code x. The reason is that
the constructor for \code x will strictly evaluate all of \code x's \code{val} fields when \code x
is created.


To model the fact that Scala field initializers are fully evaluated when the object is constructed,
we require field initializers in pDOT to be values or paths, rather than arbitrary terms. We use
the name \emph{stable term} for a value or path.

This raises the question of how to model a Scala method such as \code b.
A method can still be represented by making the delayed evaluation of the body explicit:
instead of initializing the field \code b with the method body itself, we delay the body inside
a lambda abstraction. The lambda abstraction, a value, can be assigned to the field \code b. The
body of the lambda abstraction can be an arbitrary term; it is not evaluated during object construction,
but later when the method is called and the lambda is applied to some dummy argument.

\subsection{Precise Self Types}\label{sec:typerestrict}

DOT allows powerful type abstractions, but it demands \textit{objects} as proof that the type abstractions make sense.
An object assigns actual types to its type members and thus provides concrete evidence for the declared
subtyping relationships between abstract type members.
To make the connection between the object value and the type system, DOT requires the self type
in an object literal to precisely describe the concrete types assigned to type members, and we
need to define similar requirements for self types in pDOT.

In the object
$\tNew x {\tTypeDec A T T} \set{ A = T }$,
DOT requires the self-type to be $\tTypeDec A T T$ rather than some wider type $\tTypeDec A S U$.
This is not merely a convenience, but it is essential for soundness. Without the requirement,
DOT could create and type the object
$\tNew x {\tTypeDec A \top \bot} {\set{A = T}}$,
which introduces the subtyping relationship $\top<:\bot$ and thus makes every type a subtype of every other
type.
Although we can require the actual assigned type~$T$
to respect the bounds (i.e. $\top <: T <: \bot$), such a condition is
not sufficient to prohibit this object.
The assigned
type $T$ and the bounds ($\top$ and $\bot$ in this example) can in general
depend on the self variable, and thus the condition makes
sense only in a typing context that contains the self variable
with its declared self type. But in such a context, we already
have the assumption that $\top <: x.A <: \bot$, so it
holds that $\top <: T$ (since $\top <: x.A <: \bot <: T$) and similarly
$T <: \bot$.



In pDOT, a
path-dependent type $p.A$ can refer to type members not only at the
top level, but also deep inside the
object.
Accordingly, we need to extend the precise self type requirement to
apply recursively within the object, as follows:
\begin{enumerate}
    \item An object containing a type member definition $\set{ A = T }$ must
    declare $A$ with tight bounds, using
$\tTypeDec A T T$ in its self type.

\item An object containing a  definition $\set{a =\tNew x T d }$ must
declare $a$ with the recursive type $\tRec x T$, using
$\tFldDec a {\tRec x T }$ in its self type.

\item An object containing a definition $\set{ a = \tLambda x T U}$ must
declare $a$ with a function type, using
$\tFldDec a {\tForall x S V}$ in its self type.

\item An object containing a definition $\set{ a = p }$ must declare $a$
with the singleton type $\single p$, using
$\tFldDec a {\single p }$ in its self type.
\end{enumerate}
The first requirement is the same as in DOT. The second and third
requirements are needed for soundness of paths that select type members
from deep within an object. The fourth requirement is needed to prevent
unsoundness in the case of cyclic references. For example, if we were to allow
the object
$\tNew x {\tFldDec a {\tTypeDec A \top \bot}} {\set{a = x.a}}$
we would again have $\top <: \bot$. The fourth requirement forces this
object to be declared with a precise self type:
$\tNew x {\tFldDec a {\single {x.a}}} {\set{a = x.a}}$.
Now, $x.a$ no longer has the type $\tTypeDec A \top \bot$, so it no longer collapses
the subtyping relation.
The precise typing thus ensures that cyclic paths can be only typed with singleton types
but not function or object types,
and therefore we cannot have type or term selection on cyclic paths.

Although both DOT and pDOT require precision in the self type of an
object, the object itself can be typed with a wider type once it is
assigned to a variable. For example, in DOT, if we have

\vspace{-2mm}
\begin{align*}
    \ttLet x {\tNew x {\tTypeDec A T T} {\set{A=T}}} {\dots}
\end{align*}
then $x$ also has the wider type $\tTypeDec A \bot \top$.
Similarly, in pDOT, if we have
\begin{align*}
    \ttLet x {\tNew x {\tFldDec a {\tRec y {\tFldDec b {\tForall z T U}} \wedge \tFldDec c {\single {x.a.b}}}} d} \dots
\end{align*}
then $x$ also has all of the following types:
$\tFldDec a {\tFldDec b {\tForall z T U}}$,
$\tFldDec a {\tRec y {\tFldDec b \top}}$,
$\tFldDec c {\single {x.a.b}}$, and
$\tFldDec c {\tForall z T U}$.
In fact, the typings for this object in pDOT are more expressive than in DOT.
Because DOT does not open types nested inside of field declarations,
DOT cannot assign the first two types to $x$.
In Section~\ref{sec:typing}, we show one simple type rule that generalizes pDOT
to open and abstract types of term members nested deeply inside an object.
In Section~\ref{sec:examples}, we encode several examples from previous DOT papers in \pdot
and show that the real-world compiler example from Section~\ref{sec:intro} that uses
types depending on long paths can be encoded in \pdot as well.

In summary, both DOT and pDOT require the self type in an object literal to precisely
describe the values in the literal, but this does not limit the ability to
ascribe a more abstract type to the paths that identify the object.



\section{From DOT to \pdot}\label{sec:pdot}

The \pdot calculus generalizes DOT by allowing paths wherever DOT allows variables
(except in places where variables are used as binders, such as $x$ in $\tLambda xTt$).

\subsection{Syntax}
\begin{wide-rules}\noindent
	{\footnotesize\begin{multicols}{3}\noindent
			\begin{flalign}
			x,\,y,\,z            \tag*{\textbf{Variable}}\\
			a,\,b,\,c            \tag*{\textbf{Term member}}\\
			A,\,B,\,C            \tag*{\textbf{Type member}}\\
			\new {p,\,q,\,r}\coloneqq\ &           \tag*{\tnew{\textbf{Path}}}\\
			&x                                      \tag*{variable}\\
			&\new{p.a}                              \tag*{field selection}\\
			t,\,u\coloneqq\ &                     \tag*{\textbf{Term}}\\
			&\assignTrm                             \tag*{stable term}\\
			&\new{p\,q}                             \tag*{application}\\
			&\tLet x t u                           \tag*{let binding}\\ \tag*{} \\ \tag*{} \\
			\assignTrm\coloneqq\ &                 \tag*{\textbf{Stable Term}}\\
			&\new p                                 \tag*{path}\\
			&v         	                            \tag*{value}\\
			v\coloneqq\ &                           \tag*{\textbf{Value}}\\
			&\tNew x T d                            \tag*{object}\\
			&\tLambda x T t                         \tag*{lambda}\\
			d\coloneqq\ &                            \tag*{\textbf{Definition}}\\
			&\set{a=\new {\assignTrm}}              \tag*{field definition}\\
			&\set{A=T}                              \tag*{type definition}\\
			&\tAnd d {d'}                           \tag*{aggregate definition}\\
			        &\tag*{}\\
			S,\,T,\,U,\,V\coloneqq &                     \tag*{\textbf{Type}}\\
			&\top                                   \tag*{top type}\\
			&\bot                                   \tag*{bottom type}\\
			&\tFldDec a T                           \tag*{field declaration}\\
			&\tTypeDec A S T                        \tag*{type declaration}\\
			&\tAnd S T                              \tag*{intersection}\\
			&\tRec x T                              \tag*{recursive type}\\
			&\tForall x S T                         \tag*{depend. function}\\
			&\new p.A                               \tag*{type projection}\\
			&\new{\single p}                        \tag*{\tnew{singleton type}}
			\end{flalign}
	\end{multicols}}
	\caption{Abstract syntax of \pdot}  
	\label{fig:synt}
\end{wide-rules}
Figure~\ref{fig:synt} shows the abstract syntax of \pdot which is based on the
DOT calculus of~\citet{wadlerfest}.
Differences from that calculus are indicated by shading.

The key construct in \pdot is a path, defined to be a variable
followed by zero or more field selections (e.g. $x.a.b.c$).
\pdot uses paths wherever DOT uses variables.
In particular, field selections $x.a$ and function application $x\,y$ are done on
paths: $p.a$ and $p\,q$.
Most importantly, \pdot also generalizes DOT's types by allowing path-dependent types~$p.A$
on paths rather than just on variables.
Additionally, as described in Section~\ref{sec:sngl}, the \pdot calculus formalizes
Scala's singleton types.
A singleton type $\single p$ is inhabited with only one value:
the value that is assigned to the path~$p$.
A singleton type thus indicates that a term designates the same object as the path~$p$.
Just as a path-dependent type~$p.A$ depends on the value of~$p$, a singleton type~$\single q$
depends on the value of $q$. Singleton types are
therefore a second form of dependent types in the calculus.

\subsection{\pdot Typing Rules}\label{sec:typing}

The typing and subtyping rules of \pdot are shown in Figures~\ref{fig:typing} and~\ref{fig:subtyping}.
The type system is based on the DOT of~\citet{wadlerfest}, and all changes are
highlighted in gray.
\begin{wide-rules}\footnotesize
\begin{flalign*}
  \G&\Coloneqq \varnothing\ |\ \G,\,x\colon T
                       \tag*{Type environment}
\end{flalign*}

\textbf{Term typing}
\begin{multicols}{3}

\infrule[Var]
  {\G(x)=T}
  {\typDft x T}

\infrule[All-I]
{\typ {\extendG x T} t U
	\andalso
	x\notin\fv T}
{\typDft{\tLambda x T t}{\tForall x T U}}

\infrule[All-E]
{\typDft {\new p} {\tForall z S T}
	\andalso
	\typDft {\new q} S}
{\typDft {\new{p\, q}} {\tSubst z {\new q} T}}
  
\infrule[\{\}-I]
{\dtypN x {\extendG x T} d T}
{\typDft {\tNew x T d} {\tRec x T}}

\infrule[Fld-E]
{\typDft {\new p} {\tFldDec a T}}
{\typDft {\new p.a} T}
  
\newruletrue

\infrule[Fld-I]
{\typDft {p.a} T}
{\typDft {p} {\tFldDec a T}}

\newrulefalse

\infrule[Let]
{\typDft t T
	\\
	\typ {\extendG x T} u U
	\andalso
	x\notin\fv U}
{\typDft {\ttLet x t u} U}

\newruletrue

\infrule[Sngl-Trans]
	{\typDft p {\single q} \andalso \typDft q T}
	{\typDft p T}

\newruletrue
\infrule[Sngl-E]
  {\typDft p {\single q} \andalso
   \typeable {q.a}}
  {\typDft {p.a} {\single {q.a}}}
\newrulefalse

\infrule[Rec-I]
  {\typDft {\new p} {\tSubst x {\new p} T}}
  {\typDft {\new p} {\tRec x T}}

\infrule[Rec-E]
  {\typDft {\new p} {\tRec x T}}
  {\typDft {\new p} {\tSubst x {\new p} T}}

\infrule[\&-I]
  {\typDft {\new p} T
    \andalso
    \typDft {\new p} U}
  {\typDft {\new p} {\tAnd T U}}

\infrule[Sub]
  {\typDft t T
    \andalso
    \subDft T U}
  {\typDft t U}

\end{multicols}

\textbf{Definition typing}
\begin{multicols}{2}

 \infax[Def-Typ]
 {\dtypDftN {\set{A=T}} {\tTypeDec A T T}}  
 
\newruletrue
\infrule[Def-All]
  {\typDft {\tLambda x T t} {\tForall x U V}}
  {\dtypDft {\set{a=\tLambda x T t}} {\tFldDec a {\tForall x U V}}}

\infrule[Def-New]
  {\dtyp {p.a} \G {\tSubst y {p.a} d} {\tSubst y {p.a} T}
   \andalso
   \tightboundseq T}
  {\dtypDft {\set{a = \tNew y T d}} {\tFldDec a {\tRec y T}}}

\infrule[Def-Path]
  {\typeable q}
  {\dtypDft {\set{a = q}} {\tFldDec a {\single q}}}
  
\newrulefalse
  
\infrule[AndDef-I]
  {\dtypDftN {d_1} {T_1}
    \andalso
    \dtypDftN {d_1} {T_2}
    \\
    \dom{d_1},\,\dom{d_2}\text{ disjoint}}
  {\dtypDftN {\tAnd {d_1} {d_2}} {\tAnd {T_1} {T_2}}}
\end{multicols}
\begin{multicols}{2}
\noindent
\textbf{Typeable paths}
\newruletrue
\infrule[Wf]
    {\typDft p T}
    {\typeable p}
\newrulefalse
\noindent
\textbf{Tight bounds}
\[
\new{\tightboundseq T} =
\begin{cases}
U = V							&\text{if } T = \tTypeDec A U V\\
\tightboundseq U         			&\text{if } T = \tRec x U \text{ or } \tFldDec a U\\
\tightboundseq U \text{ and }
    \tightboundseq V           	&\text{if } T = U \wedge V\\
\text{true}           			&\text{otherwise}
\end{cases}
\]
\end{multicols}	
\caption{\pdot typing rules}
\label{fig:typing}
\end{wide-rules}
\begin{wide-rules}
  \begin{multicols}{3}
\infax[Top]
  {\subDft T \top}

\infax[Bot]
  {\subDft \bot T}

\infax[Refl]
  {\subDft T T}

\infrule[Trans]
  {\subDft S T
    \andalso
    \subDft T U}
  {\subDft S U}

\infax[And$_1$-$<:$]
  {\subDft {\tAnd T U} T}

\infax[And$_2$-$<:$]
  {\subDft {\tAnd T U} U}

\infrule[$<:$-And]
  {\subDft S T
    \andalso
    \subDft S U}
  {\subDft S {\tAnd T U}}
  
\infrule[Fld-$<:$-Fld]
{\subDft T U}
{\subDft {\tFldDec a T} {\tFldDec a U}}

\infrule[Typ-$<:$-Typ]
{\subDft {S_2} {S_1}
	\andalso
	\subDft {T_1} {T_2}}
{\subDft {\tTypeDec A {S_1} {T_1}} {\tTypeDec A {S_2} {T_2}}}

\infrule[$<:$-Sel]
{\typDft {\new p} {\tTypeDec A S T}}
{\subDft S {{\new p}.A}}

\newruletrue
\infrule[Sngl$_{pq}$-$<:$]
{\typDft p {\single q} \andalso \typeable q}
{\subDft T {\repl p q T}}

\infrule[Sngl$_{qp}$-$<:$]
{\typDft p {\single q} \andalso \typeable q}
{\subDft T {\repl q p T}}

\newrulefalse

\infrule[Sel-$<:$]
  {\typDft {\new p} {\tTypeDec A S T}}
  {\subDft {{\new p}.A} T}

\infrule[All-$<:$-All]
{\subDft {S_2} {S_1}
	\\
	\sub {\extendG x {S_2}} {T_1} {T_2}}
{\subDft {\tForall x {S_1} {T_1}} {\tForall x {S_2} {T_2}}}
\normalsize
  \end{multicols}
\caption{\pdot subtyping rules}
\label{fig:subtyping}
\end{wide-rules}
\subsubsection{From Variables to Paths}

The first thing to notice in the \pdot typing and subtyping rules is
that all variable-specific rules, except \rn{Var}, are generalized to paths,
as motivated in Section~\ref{sec:pathsvar}.
The key rules that make DOT and \pdot
interesting are the type-selection rules
\rn{<:-Sel} and \rn{Sel-<:}. These rules enable
us to make use of the type member in a path-dependent type.
When a path $p$ has type $\tTypeDec A S U$, the rules introduce
the path-dependent type $p.A$ into the subtyping relation by
declaring the subtyping constraints $S <: p.A$ and $p.A <: U$.
Thanks to these two rules, \pdot supports fully path-dependent types.

\subsubsection{Object Typing}
Similarly to the DOT calculus, the \rn{\{\}-I} rule gives an object $\tNew x T d$ with declared
type $T$ which may depend on the self variable $x$ the recursive type
$\tRec x T$. The rule also checks that the definitions $d$ of the object
actually do have type $T$ under the assumption that the self variable
has this type. The object's definitions~$d$ are checked by the Definition typing rules.
As discussed in Section~\ref{sec:typerestrict},
the rules assign a precise self type for objects, ensuring that
paths are declared with singleton types,
functions with function types,
and objects with object types.
For objects, the $\tightbounds{T}$ condition ensures that all type members that
can be reached by traversing $T$'s fields have equal bounds,
while still allowing arbitrary bounds in function types.

A difference with DOT is that \pdot's definition-typing judgment keeps track
of the path that represents an object's identity.
When we typecheck an outermost object that
is not nested in any other object, we use the \rn{\{\}-I} rule.
The rule introduces $x$ as the identity for the object
and registers this fact in the definition-typing judgment.
To typecheck an object that is assigned to a field $a$ of another object $p$ we use the \rn{Def-New} rule.
This rule typechecks the object's body assuming the object identity $p.a$ and
replaces the self-variable of the object with that path.
The definition-typing judgment keeps track of the path to the definition's
enclosing object starting
from the root of the program. This way the type system knows what
identities to assign to nested objects. For example, when typechecking the
object assigned to $x.a$ in the expression
$$
    \tLet x {\nu(x)\set{a=\nu(y)\set{b=y.b}}} \dots
$$
we need to replace $y$ with the path $x.a$:

\vspace{-2mm}
\begin{align*}
    \inferrule*[Right=\parbox{2cm}{\rn{{\tiny \{\}-I}}}]
        {\inferrule*[Right=\parbox{2cm}{\rn{{\tiny Def-New}}}]
            {\inferrule*[Right=\parbox{2cm}{\rn{{\tiny Def-Path}}}]
                {\extendG x {\tFldDec a {\tRec y {\tFldDec b 
                                {\single{y.b}}}}}
                            \vdash x.a}
                {\dtyp {\hleq{x.a}} {\extendG x {\tFldDec a {\tRec y {\tFldDec b 
                    {\single{y.b}}}}}}
                {\set{b=\hleq{x.a}.b}}
                {\tFldDec b {\single{\hleq{x.a}.b}}}}
            \qquad\quad
            \tightboundseq {\tFldDec b {\single{y.b}}}}
            {\dtyp {\hleq x} {\extendG x {\tFldDec a {\tRec y {\tFldDec b {\single{y.b}}}}}}
                 {\set{a=\nu(y)\set{b=y.b}}}
                 {\tFldDec a {\tRec y {\tFldDec b {\single{y.b}}}}}}}
        {\typDft {\nu(x)\set{a=\nu(y)\set{b=y.b}}} 
                 {\tRec x {\tFldDec a {\tRec y {\tFldDec b {\single{y.b}}}}}}}
\end{align*}
An alternative design of the \rn{Def-New} rule
can be to introduce a fresh variable $y$ into the 
context (similarly to the \rn{\{\}-I} rule). However, we would have to assign $y$
the type $\single{x.a}$ to register the fact that these two paths identify the same object.
We decided to simplify the rule by immediately replacing the nested object's self
variable with the outer path to avoid the indirection of an additional singleton type.

\subsubsection{Path Alias Typing}

In \pdot singleton-type related typing and subtyping rules are responsible for the handling of aliased paths and equivalent types.

\paragraph{Singleton Type Creation}
How does a path $p$ obtain a singleton type?
A singleton type indicates that in the initial program, a prefix of $p$ (which could be all of $p$) is assigned a path $q$. 
For example, in the program
$$
    \tLet x {\tNew x {\tFldDec a {\single x} \wedge \tFldDec b S}\set{a=x}\wedge\set{b=s}} {\dots}
$$
the path $x.a$ should have the type $\single x$ because $x.a$ is assigned the path $x$.
The singleton type for $x.a$ can be obtained as follows. Suppose that in the typing context of the \mlet body, $x$ is mapped to
the type of its object,
$\tRec x {\tFldDec a {\single x} \wedge \tFldDec b S}$. Through applying recursion elimination (\rn{Rec-E}),
field selection (\rn{Fld-E}), and finally subsumption (\rn{Sub}) with the intersection subtyping rule \rn{And$_1$-$<:$},
we will obtain that $\typDft {x.a} {\single x}$.

In the above example, $x.a$ aliases $x$, so anything that we can do with $x$ we should be able to do with $x.a$.
Since $x$ has a field $b$ and we can create a path $x.b$,
we want to be also able to create a path $x.a.b$. Moreover, we want to treat $x.a.b$ as an alias for $x.b$.
This is done through the \rn{Sngl-E} rule: it says that if $p$ aliases $q$, and
$q.a$ has a type (denoted with $\typeable {q.a}$), then $p.a$ aliases $q.a$.
This rule allows us to conclude that $\typDft {x.a.b} {\single{x.b}}$.

\paragraph{Singleton Type Propagation}

In the above example we established that the path $x.a.b$ is an alias for $x.b$.
Therefore, we want to be able to type $x.a.b$ with any type with which we can type $x.b$.
The \rn{Sngl-Trans} rule allows us to do just that: if $p$ is an alias for $q$, then we can type $p$
with any type with which we can type $q$. Using that rule, we can establish that $\typDft {x.a.b} S$
because $\typDft {x.b} S$.

\paragraph{Equivalent Types}\label{sec:equivtypes}

\newcommand{\ppath}{\hleq p}
\newcommand{\rpath}{{\hleqq q}}

As described in Section~\ref{sec:replacement}, we call two types equivalent if they are equal up to path aliases.
We need to ensure that equivalent types are equivalent by subtyping, i.e.~that they are subtypes of each other.
For example, suppose that $\typDft \ppath {\single \rpath}$, and the path $r$ refers to an object
$\nu(x)\set{a=\ppath}\wedge\set{b=\ppath}$. Then we want to be able to type $r$ with all of
the following types: 

\begin{align*}
&\typDft r {\tFldDec a {\single {\ppath }} \wedge \tFldDec b {\single {\ppath }}} \quad 
&&\typDft r {\tFldDec a {\single {\ppath }} \wedge \tFldDec b {\single {\rpath}}} \\
&\typDft r {\tFldDec a {\single {\rpath}} \wedge \tFldDec b {\single {\rpath}}} \quad
&&\typDft r {\tFldDec a {\single {\rpath}} \wedge \tFldDec b {\single {\ppath }}}
\end{align*}
\noindent
The \pdot subtyping rules \rn{Sngl$_{pq}$-$<:$} and \rn{Sngl$_{qp}$-$<:$} allow us to
assign these types to $r$ by establishing subtyping between equivalent types.
Specifically, if we know that $\typDft p {\single q}$ then the rules allow us to replace
any occurrence of $p$ in a type $T$ with $q$, and vice versa, while maintaining subtyping relationships in both directions.

We express that two types are equivalent using the \textit{replacement operation}.
The operation is similar to the substitution operation, except that we replace
paths with paths instead of variables with terms, and we replace only one path at a time rather than all of its occurrences.
The statement $\repl p q T = U$ denotes that the type $T$ contains
one or more paths that start with $p$, e.g. $p.\overline{b_1}$, \dots, $p.\overline{b_n}$,
and that exactly one of these occurrences $p.\overline{b_i}$
is replaced with $q.\overline {b_i}$,
yielding the type $U$.
Note that it is not specified exactly in which occurrence of the above paths
the prefix $p$ is replaced with $q$.
The precise definition of the replacement operation
is presented in Figure~\ref{fig:repl}.
In the figure, the notation $p.\overline b$ denotes
a path with a possibly empty sequence of field selections, e.g.~$p.b_1.b_2$.

\begin{wide-rules}
\begin{multicols}{2}

\newruletrue

\infax[Repl-Path]
  {\repl p q {p.\overline b.A} = q.\overline b.A}

\infax[Repl-Sngl]
  {\repl p q {\single {p.\overline b}} = \single {q.\overline b}}
  
\infrule[Repl-And$_1$]
  {\repl p q {T_1} = T_2}
  {\repl p q {(T_1 \wedge U)} = T_2 \wedge U}
  
\infrule[Repl-And$_2$]
  {\repl p q {T_1} = T_2}
  {\repl p q {(U\wedge T_1)} = U\wedge T_2}
   
\infrule[Repl-Rec]
  {\repl p q T = U}
  {\repl p q {\tRec x T} = \tRec x U}

\infrule[Repl-All$_1$]
 {\repl p q {T_1} = T_2}
 {\repl p q {(\tForall x  {T_1} U)} = \tForall x {T_2} U}
 
\infrule[Repl-All$_2$]
	{\repl p q {T_1} = T_2}
 	{\repl p q  {(\tForall x U {T_1})} = \tForall x U {T_2}}
  
\infrule[Repl-Fld]
  {\repl p q {T_1} = T_2}
  {\repl p q {\tFldDec a {T_1}} = \tFldDec a {T_2}}
  
\infrule[Repl-Typ$_1$]
 {\repl p q {T_1} = T_2}
 {\repl p q {\tTypeDec A {T_1} U} = \tTypeDec A {T_2} U}
 
\infrule[Repl-Typ$_2$]
 {\repl p q {T_1} = T_2}
 {\repl p q {\tTypeDec A U {T_1}} = \tTypeDec A U {T_2}}\
 
 \newrulefalse

\end{multicols}
\caption{Replacement of a path $p$ in a type by $q$}
\label{fig:repl}
\end{wide-rules}

Given the path $r$ from the above example, we can choose whether to replace the first or second occurrence of $\ppath$ with $\rpath$; for example, we can derive

\vspace{-4mm}
\begin{align*}
{\tiny
    \inferrule*[Right=\parbox{2cm}{\rn{{\tiny Sub}}}]
        {\inferrule*[Right=\parbox{2cm}{\rn{{\tiny Rec-E}}}]
            {\dots}{\G \vdash r \colon \tFldDec a {\single {\ppath }} \wedge \tFldDec b {\single {\ppath }}}\hspace{1cm}
         \inferrule*[Right=\parbox{2cm}{\rn{{\tiny\hleq{Sngl$_{pq}$-$<:$}}}}]
         {\typDft \ppath {\single \rpath} \quad 
            \inferrule*[Right=\parbox{2cm}{\rn{{\tiny Repl-And$_2$}}}]
            {\dots}
            {\repl \ppath \rpath
                       {\tFldDec a {\single {\ppath }} \wedge \tFldDec b {\single {\ppath }}}=
                       {\tFldDec a {\single {\ppath }} \wedge \tFldDec b {\single {\rpath}}}}}
        {\G \vdash \tFldDec a {\single {\ppath }} \wedge \tFldDec b {\single {\ppath }} <:
            \tFldDec a {\single {\ppath }} \wedge \tFldDec b {\single {\rpath}}}}
    {\G \vdash r \colon \tFldDec a {\single {\ppath }} \wedge \tFldDec b {\single {\rpath}}}}
\end{align*}
To replace several occurrences of a path with another, we repeatedly apply \rn{Sngl$_{pq}$-$<:$} or \rn{Sngl$_{qp}$-$<:$}.

\subsubsection{Abstracting Over Field Types}\label{sec:abstractfield}

Finally, we describe one of the most interesting \pdot rules which adds significant expressivity to \pdot.

Consider a function
$f=\tLambda x {\tFldDec a T} {\dots}$
and a path $p$ that refers to the object $\nu(x\colon \tFldDec a {\single{q}})$ $\set{a=q}$,
where $\typDft q T$.
Since $\typDft p {\tRec x {\tFldDec a {\single q}}}$, by \rn{Rec-E},
$\typDft p {\tFldDec a {\single q}}$, assuming that $q$ does not start with $x$.
Therefore, since $\typDft q T$, we would like to be able to pass $p$ into the function~$f$
which expects an argument of type $\tFldDec a T$.
Unfortunately, the typing rules so far do not allow us to do that because although
$q$ has type $T$, $\single q$ is \textit{not} a subtype of $T$, and therefore
$\tFldDec a {\single {q}}$ is not a subtype of $\tFldDec a T$.

The type rule \rn{Fld-I} allows us to bypass that limitation. If a path $p$ has a record type $\tFldDec a T$
(and therefore $\typDft {p.a} T$),
then the rule lets us type $p$ with any type $\tFldDec a U$ as long as $p.a$ can be typed with $U$.

For the above example, we can prove that $\typDft p {\tFldDec a T}$ and pass it into $f$ as follows:

\vspace{-2mm}
\begin{align*}
    \inferrule*[Right=\parbox{2cm}{{\scriptsize \hleq{\rn{Fld-I}}}}]
    {
        \inferrule*[Right=\parbox{2cm}{{\scriptsize \rn{Sngl-Trans}}}]
            {\inferrule*[Right=\parbox{2cm}{{\scriptsize \rn{Fld-E}}}]
                {\typDft p {\tFldDec a {\single q}}}
                {\typDft {p.a} {\single q}}
            \qquad\qquad
            \typDft q T
            }
            {\typDft {p.a} T}
    }
    {\typDft p {\tFldDec a T}}
\end{align*}

The \rn{Fld-I} rule allows us to eliminate recursion on types that are nested inside fields, which is not possible in DOT.
If a DOT function~$f$ expects a parameter of type $\tFldDec a {\tRec x {T}}$, then in DOT,
we cannot pass a variable $y$ of type $\tFldDec{a}{\tRec x {T\wedge U}}$ or a variable $z$ of type
$\tFldDec{a}{\tSubst x {z.a} {T}}$ into~$f$
because there is no subtyping between recursive types, and there is no subtyping relationship between $\tRec x T$ and $\tSubst x {z.a} T$
(and the latter type might not exist in the first place due to the lack of fully path-dependent types).
All of the above is possible in \pdot because both $y.a$ and $z.a$ can be typed with $\tRec x T$, which allows us to use the
\rn{Fld-I} rule and type $y$ and $z$ as $\tFldDec a {\tRec x {T}}$.

\subsection{Reduction Semantics}\label{sec:reduction}
\setlength{\multicolsep}{9pt}
\begin{wide-rules}
\begin{multicols}{2}\noindent
\textcolor{gray}{
    \infrule[\redDotProj]
    {\sta(x)=\tNew x T {\dots\set{a=t}\dots}}
    {\reduction{\sta}{x.a}{\sta}{t}}}

\begin{align*}
  \sta&\Coloneqq \varnothing\ |\ \sta\co x\mapsto v
                       \tag*{\textbf{Store}}\\
\end{align*}
\end{multicols}
\begin{multicols*}{2}
{\color{gray}
\infrule[\redDotApply]
    {\matchnewhight{\gamma(x)=\tLambda z T t}}
    {\reductionDft
        {\matchnewhight{x\,y}}
        {\tSubst z y t}}}
\infrule[\redApply]
{\new{\lookupDft p {\tLambda z T t}}}
{\reductionDft
    {\new{p\,q}}
    {\tSubst z {\new q} t}}
\end{multicols*}
\begin{multicols*}{2}
{\color{gray}
\infax[\redDotLetvar]
{\reductionDft
    {\ttLet x y t}
    {\tSubst x y t}}}
\infax[\redLetpath]
{\reductionDft
    {\ttLet x {\new p} t}
    {\tSubst x {\new p} t}}
\end{multicols*}
\begin{multicols*}{2}
{\color{gray}
\infrule[\redDotLetvalue]
    {x \notin \dom\sta}
    {\reduction
        \sta
        {\ttLet x v t}
        {\extendSta x v} 
        t}}
\infrule[\redLetvalue]
{x \notin \dom\sta}
{\reduction
    \sta
    {\ttLet x v t}
    {\extendSta x v} 
    t}
\end{multicols*}
\begin{multicols*}{2}
{\color{gray}
\infrule[\redDotCtx]
    {\reduction
        \sta
        t
        {\sta'}{t'}}
    {\reduction
        \sta{\ttLet x t u}
        {\sta'}{\ttLet x {t'} u}}}
\infrule[\redCtx]
{\reduction
    \sta
    t
    {\sta'}{t'}}
{\reduction
    \sta{\ttLet x t u}
    {\sta'}{\ttLet x {t'} u}}
\end{multicols*}

\caption{Operational semantics of \textcolor{gray}{DOT} and \pdot}
\label{fig:red2}
\end{wide-rules}

\begin{wide-rules}
\begin{multicols}{3}
\newruletrue

	\infrule[Lookup-Step-Var]
		{\sta(x) = v}
		{\lookupStepDft x v}

	\infrule[Lookup-Step-Val]
		{\lookupStepDft p {\tNew x T {\dots \set{a=\assignTrm} \dots}}}
		{\lookupStepDft {p.a} {\tSubst x p \assignTrm}}		
		
	\infrule[Lookup-Step-Path]
		{\lookupStepDft p q}
		{\lookupStepDft{p.a}{q.a}}
\end{multicols}
\begin{multicols}{2}
\newruletrue
	\infax[Lookup-Refl]
        {\lookupDft s s}
    
   	\infrule[Lookup-Trans]
        {\lookupStepDft {s_1} {s_2} \andalso \lookupDft {s_2} {s_3}}
        {\lookupDft {s_1} {s_3}}
		
\newrulefalse
\end{multicols}

\caption{Value-environment path lookup}
\label{fig:lookup}
\end{wide-rules}
\setlength{\multicolsep}{2pt}
The operational semantics of \pdot is presented in Figure~\ref{fig:red2}. 
\pdot's reduction rules mirror the DOT rules with three distinctions:
\begin{itemize}
    \item \textit{paths everywhere}: wherever DOT uses (as opposed to defines) variables, \pdot uses paths;
    \item \textit{no {\sc\redDotProj}}: there is no reduction rule for field projection because in \pdot, paths are normal form
    	(as motivated in Section~\ref{sec:pathsnormalform});
    \item \textit{path lookup}: \pdot uses the reflexive, transitive closure of the \textit{path lookup} operation $\lookupSymbol$ that generalizes 
    variable lookup in value environments to paths.
\end{itemize}
The path lookup operation is presented in Figure~\ref{fig:lookup}. This operation allows us to look up a value that is nested
deeply inside an object. If a path is a variable the lookup operation is a straightforward variable lookup (\rn{Lookup-Step-Var}).
If in a value environment $\sta$, a path $p$ is assigned an object $\nu(x)\set{a=s}$ then 
$\lookupStepDft {p.a} \tSubst x p s$  because the self variable $x$ in $s$ gets replaced with $p$ (\rn{Lookup-Step-Val}).
If $p$ is equal to another path $q$ then $\lookupStepDft {p.a}{q.a}$  (\rn{Lookup-Step-Path}).

Finally, we want to be able to follow a sequence of paths in a value environment:
for example, if $\lookupStepDft p q$ and $\lookupStepDft q v$, we want to conclude that looking up $p$ yields $v$.
This is done through the reflexive, transitive closure $\lookupSymbol^*$ of the $\lookupSymbol$ relation (\rn{Lookup-Refl} and \rn{Lookup-Trans}).

For example, looking up $x.a.c$ in the environment 
$\sta=(y\co \nu(y') \{ b = \nu(y'') \{ c = \tLambda z \top z \} \}))$,
$(x\co \nu(x) \{ a = y.b \})$
yields $\tLambda z \top z$:

\vspace{-2mm}
\newcommand{\rmsp}{\hspace{-8mm}}
\newcommand{\rmsph}{\hspace{-3mm}}
\begin{align*}
&\sta (x)&&\rmsp=   &&\rmsp &&\rmsp \nu(x) \{ &&\rmsp a = &&\rmsp y.b \}\qquad\quad
&& \sta (y)&&\rmsp=   &&&&\rmsp\rmsph \nu(y') \{ &&\rmsp b =             &&\rmsp \nu(y'') \{ &&\rmsp c =             &&\rmsp \tLambda z \top z \} \}\\
&\sta &&\rmsp\vdash &&\rmsp x \lookupSymbol &&\rmsp \nu(x) \{ &&\rmsp a = &&\rmsp y.b\}
&& \sta &&\rmsp\vdash &&\rmsp y \lookupSymbol &&\rmsp\rmsph \nu(y') \{ &&\rmsp b =             &&\rmsp \nu(y'') \{ &&\rmsp c =             &&\rmsp \tLambda z \top z \} \}\\
&\sta &&\rmsp\vdash &&\rmsp x.              &&\rmsp           &&\rmsp a \lookupSymbol &&\rmsp y.b
&& \sta &&\rmsp\vdash &&\rmsp y.              &&\rmsp           &&\rmsp b \lookupSymbol &&\rmsp \nu(y'') \{  &&\rmsp c =             &&\rmsp \tLambda z \top z \}\\
&\sta &&\rmsp\vdash &&\rmsp x.              &&\rmsp           &&\rmsp a.c\lookupSymbol&&\rmsp y.b.c
&& \sta &&\rmsp\vdash &&\rmsp y.              &&\rmsp           &&\rmsp b.              &&\rmsp              &&\rmsp c \lookupSymbol &&\rmsp \tLambda z \top z
\end{align*}
\\[-8mm]
\begin{align*}
\hspace{-33mm}\lookupDft {x.a.c} {\tLambda z \top z}
\end{align*}

The reduction rule that uses the lookup operation is the function application rule \rn{Apply}:
to apply $p$ to $q$ we must be able to look up $p$ in the value environment and obtain
a function.
Since \pdot permits cycles in paths, does this mean that the
lookup operation for this type rule might not terminate?
Fortunately, \pdot's type safety ensures that this will not happen.
As shown in Section~\ref{sec:lookupterminates}, if $\typDft p \tForall T U$ then lookup of~$p$
eventually terminates and results in a function value. Therefore, a function application $p\,q$ always
makes progress.

\section{Type Safety}\label{sec:proof}

We implemented the type-safety proof of \pdot in Coq as an extension of the simple DOT soundness proof by~\citet{simpl}.
Compared to the 2,051 LOC, 124 lemmas and theorems, and 65 inductive or function
definitions in the simple DOT proof, the \pdot Coq formalization consists of
7,343 LOC, 429 lemmas and theorems, and 115 inductive or function definitions.
Our paper comes with an artifact that presents the Coq formalization.
A correspondence between the presentation of \pdot and the proof in the paper,
and the Coq mechanization is presented in Section~\ref{sec:correspondence}.
This section presents an overview of the key challenges and insights of proving
pDOT sound.

Type safety ensures that any well-typed \pdot program does not get stuck,
i.e. it either diverges or reduces to a normal form (a path or a value):

\begin{theorem}[Type Soundness]\label{theorem}
    If\ $\typ {} t T$ then either $t$ diverges, i.e. there exists an infinite reduction sequence
    $\st \varnothing t\red \st{\sta_1}{t_1}\red \dots\red \st{\sta_n}{t_n}\red \dots$ starting with $t$, or
    $t$ reduces to a normal form~$s$, i.e. $\st\varnothing t\redt \st\sta s$,
    and $\typDft s T$ for some $\G$ such that $\wfDft$.
\end{theorem}

Since evaluating \pdot programs can result in paths (which are normal form),
one might ask whether looking up those paths yields anything meaningful.
As mentioned in Section~\ref{sec:welltypedpaths}, 
looking up any well-typed path in the runtime environment results
either in a value or an infinite loop.
To formulate the final soundness theorem that reasons about both
term reduction and path lookup we define the following
extended reduction relation~$\redext$:

\begin{multicols}{2}
    \infrule
    {\reduction \sta t {\sta'} {t'}}
    {\stDft t \redext {\st {\sta'} {t'}}}
    \infrule
    {\lookupStep \sta s {\sta'} {s'}}
    {\stDft s \redext {\st {\sta'} {s'}}}
\end{multicols}

We denote the reflexive, transitive closure of extended
reduction ad $\redextt$. Finally,
we state the following extended soundness theorem:
\begin{theorem}[Extended Type Soundness]\label{theorem-extended}
    If\ $\typ {} t T$ then either $t$ diverges, i.e. there exists an infinite reduction sequence
    $\st \varnothing t\redext \st{\sta_1}{t_1}\redext \dots\redext \st{\sta_n}{t_n}\redext \dots$ starting with $t$, or
    $t$ reduces to a \textit{value}, i.e. $\st\varnothing t\redextt \st\sta v$.
\end{theorem}

\noindent
Our proof follows the syntax-based approach to type soundness by~\citet{felleisen}. The two central
lemmas of the proof are Progress and Preservation:

\begin{lemma}[Progress]\label{lemma:progress}
    Let $\sta$ be a value environment and $\G$ a typing environment. If
        i) $\wfDft$ (i.e. if $\sta=\overline{(x_i\co v_i)}$ then $\G=\overline{(x_i\co T_i)}$
            and $\typDft {v_i} {T_i}$),
        ii) $\G$ is \emph{inert} (i.e. all types in $\G$ are the precise types of values, see Section~\ref{sec:proofrecipe}),
        iii) \tnew{$\G$ is \emph{well-formed}} (i.e. all paths in the types of $\G$ are typeable in $\G$, see Section~\ref{sec:inertness}), and
        iv) $\typDft t T$,
    then $t$ is in normal form or there exists a term $t'$ and a value environment $\sta'$ such that
    $\reduction{\sta}{t}{\sta'}{t'}$.
\end{lemma}

\begin{lemma}[Preservation]\label{lemma:preservation}
    Let $\sta$ be a value environment and $\G$ a typing environment. If 
        i) $\wfDft$,
        ii) $\G$ is inert,
        iii) \tnew{$\G$ is well-formed},
        iv) $\typDft t T$, and
        v) $\reduction{\sta}{t}{\sta'}{t'}$,
    then there exists an inert, \tnew{well-formed} typing environment $\G'$ such that $\wf{\G'}{\sta'}$
    and $\typ{\G'} {t'} T$.
\end{lemma}

The \pdot proof follows the design principles laid out by~\citet{simpl}
of separating the reasoning about types, variables (paths), and values
from each other to ensure modularity and facilitate future extensions of DOT.

\subsection{Main Ideas of the DOT Safety Proof}\label{sec:dotsafety}
The DOT type-safety proof addresses two main challenges:
\begin{enumerate}[1)]
    \item \textit{Rule out bad bounds:}
        Although bad bounds give rise to DOT typing contexts in which undesirable subtyping relationships
        hold, the proof needs to show that all reachable run-time states can be typed in well-behaved
        contexts.
    \item \textit{Induction on typing:}
        The DOT typing rules are very flexible, mirroring intuitive notions about 
        which types a term ought to have. This flexibility requires rules that
        are opposites of each other and thus admit cycles in a derivation.
        The possibility of cycles impedes inductive reasoning about typing derivations.
\end{enumerate}

The existing type safety proof defines a notion of \textit{inert} types and typing
contexts, a simple syntactic property of types that rules out bad bounds. Specifically,
an inert type must be either a dependent function type or a recursive object type in
which type members have equal bounds (i.e. $\tTypeDec A T T$ rather than $\tTypeDec A S U$).
Crucially, the preservation lemma shows that reduction preserves inertness: that is,
when $\reduction \gamma t {\gamma'} {t'}$ there is an \textit{inert} typing environment $\G'$
that corresponds to $\gamma'$ and in which $t'$ has the required type.

The proof also employs the \textit{proof recipe}, a stratification of the typing
rules into multiple typing relations that rule out cycles in a typing derivation,
but are provably as expressive as the general typing relation under the condition
of an inert typing context. In particular, besides the general typing relation, the proof uses
three intermediate relations: \textit{tight} typing neutralizes the $<:$-\rn{Sel} and \rn{Sel}-$<:$ rules that
could introduce bad bounds, \textit{invertible} typing contains introduction rules that
create more complex types out of smaller ones, and \textit{precise} typing contains
elimination rules that decompose a type into its constituents.

\subsection{Type Safety: From DOT to \pdot}

The challenges of adapting the DOT soundness proof to \pdot can be classified
into three main themes:
adapting the notion of inert types to \pdot,
adapting the stratification of typing rules to \pdot, and
adapting the canonical forms lemma to changes in the operational semantics in \pdot.

\subsubsection{Inert Types in \pdot} \label{sec:inertness}
The purpose of inertness is to prevent the introduction of a possibly
undesirable subtyping relationship between arbitrary types $S <: U$
arising from the existence of a type member that has those types as
bounds. If a variable $x$ has type $\tTypeDec A S U$, then
$S <: x.A$ and $x.A <: U$, so by transitivity, $S <: U$.

A DOT type is \textit{inert} if it is a function type or a recursive
type $\tRec x T$
where $T$ is a \textit{record type}. A record type is either a type-member declaration
with equal bounds $\tTypeDec A U U$ or an arbitrary field declaration $\tFldDec a S$.
In DOT, this is sufficient to rule out the introduction of new subtyping relationships.

In \pdot, a new subtyping relationship $S <: U$ arises when a {\em path} $p$,
rather than only a variable~$x$, has a type member $\tTypeDec A S U$.
Therefore, the inertness condition needs to enforce equal bounds on type members
not only at the top level of an object, but recursively in any objects nested deeply
within the outermost object. Therefore, as shown in the inertness definition in Figure~\ref{fig:inert}, a field declaration $\tFldDec a T$
is inert only if the field type $T$ is also inert.
Moreover, since \pdot adds singleton types to DOT, the definition of a
record type is also extended to allow a field to have a singleton~type.
\begin{wide-rules}
\begin{multicols}{5}
\infax{\inert{\tForall x T U}}
\infrule
    {\record{T}}
    {\inert{\tRec x T}}
\infax{\record{\tTypeDec A T T}}
\newruletrue
\infax{\record{\tFldDec a {\single q}}}
\newrulefalse
\infrule
    {\new{\inert T}}
    {\record{\tFldDec a T}}
\end{multicols}
  \caption{Inert Types in \pdot}
  \label{fig:inert}
\end{wide-rules}
Both the DOT and \pdot preservation lemmas must ensure that reduction
preserves inertness of typing contexts, but \pdot also requires
preservation
of a second property, well-formedness of typing contexts.
A \pdot type can depend on paths rather
than only variables, and the type makes sense only if the paths within it
make sense; more precisely, well-formedness requires that any path that
appears in a type should itself also be typeable in the same typing
context. Without this property, it would be possible for the typing
rules to derive types for ill-formed paths, and there could be paths
that have types but do not resolve to any value during program execution.

\subsubsection{Stratifying Typing Rules in \pdot}\label{sec:proofrecipe}

The language features that \pdot adds to DOT also create new ways to introduce
cycles in a typing derivation. The stratification of the typing rules needs
to be extended to eliminate these new kinds of cycles.

The notion of aliased paths is inherently symmetric: if $p$ and $q$ are
aliases for the same object, then any term with type $p.A$ also has type
$q.A$ and vice versa. This is complicated further because the paths
$p$ and $q$ can occur deeply inside some complex type, and whether a term
has such a type should be independent of whether the type is expressed in
terms of $p$ or $q$. A further complicating factor is that a prefix of
a path is also a path, which may itself be aliased. For example, if
$p$ is an alias of $q$ and $q.a$ is an alias of $r$, then by transitivity,
$p.a$ should also be an alias of $r$.

The \pdot proof eliminates cycles due to aliased paths by breaking the symmetry.
When $p$ and $q$ are aliased, either $\typDft p {\single q}$ or $\typDft
q {\single p}$. The typing rules carefully distinguish these two cases, so that
for every pair $p, q$ of aliased paths introduced by a typing declaration,
we know whether the aliasing was introduced by the declaration of $p$ or of $q$.%
\footnote{An exceptional case is when $p$ is declared to have type $\single q$
    and $q$ is declared to have type $\single p$. Fortunately, this case of
    a cycle turns out to be harmless because neither path is declared to have
    any other type other than its singleton type, and therefore neither path
    can be used in any interesting way.
}
A key lemma then proves that if we have any sequence of aliasing relationships
$p_0 \sim p_1 \sim \dots \sim p_n$, where for each $i$, 
either $\typDft {p_i} {\single{p_{i+1}}}$ or $\typDft
{p_{i+1}} {\single{p_i}}$, we can reorder the replacements so that the ones of the
first type all come first and the ones of the second type all come afterwards.
More precisely, we can always find some ``middle'' path $q$ such that 
$\typDft{p_0}{\single q}$ and $\typDft{p_n}{\single q}$.%
\footnote{In degenerate cases, the middle path $q$ might actually be $p_0$ or $p_n$.}
Therefore, we further stratify the proof recipe into two typing judgments
the first of which accounts for the \rn{Sngl}$_{pq}$-$<:$ rule, and the second for the
\rn{Sngl}$_{qp}$-$<:$ rule.
This eliminates cycles in a typing derivation due to aliased paths,
but the replacement reordering lemma ensures that it preserves
expressiveness.

Another new kind of cycle is introduced by the field elimination rule
\rn{Fld-E} and the field introduction rule \rn{Fld-I} that is newly added in \pdot.
This cycle can be resolved in the same way as other cycles in DOT, by stratifying
these rules in two different typing relations.

The final stratification of the \pdot typing rules requires 7 typing relations
rather than the 4 required in the soundness proof for DOT.
General and tight
typing serve the same purpose as in the DOT proof, but \pdot requires
three elimination and two introduction typing relations.

\subsubsection{Canonical Forms in \pdot} \label{sec:lookupterminates}

Like many type soundness proofs, the DOT proof depends on canonical forms
lemmas that state that if a variable has a function type, then it resolves
to a corresponding function at execution time, and if it has a recursive
object type, then it resolves to a corresponding object. The change from
DOT to \pdot involves several changes to Canonical~Forms.

Two changes are implied directly by the changes to the operational
semantics. The DOT canonical forms lemmas apply to variables. Since
the \pdot \rn{Apply} reduction rule applies to paths rather than variables,
the canonical forms lemma is needed for paths. Since paths are normal
forms in \pdot and there is no \rn{Proj} reduction rule for them, on the
surface, \pdot needs a canonical forms lemma only for function types
but not for object types. However, to reason about a path with a function
type, we need to reason about the prefixes of the path, which have
an object type. Therefore, the induction hypothesis in the canonical forms lemma for
function types must still include canonical forms for object types.
Moreover, since \pdot adds singleton types to the type system, the induction hypothesis
needs to account for them as well.

A more subtle but important change is that lookup of a path in an
execution environment is a recursive operation in \pdot, and therefore
its termination cannot be taken for granted. An infinite loop in path lookup
would be a hidden violation of progress for function application, since
the APPLY reduction rule steps only once path lookup has finished finding
a value for the path.
Therefore, the canonical
forms lemma proves that if a path has a function type, then lookup
of that path does terminate, and the value with which it terminates
is a function of the required type. The intuitive argument for termination
requires connecting the execution environment with the typing environment:
if direct lookup of path $p$ yields another path $q$, then the context
assigns $p$ the singleton type $\single q$. But in order for $p$ to have
a function type, there cannot be a cycle of paths in the typing context
(because a cycle would limit $p$ to have only singleton types), and therefore
there cannot be a cycle in the execution environment.
The statement of the canonical forms lemma is:
\begin{lemma}\label{lem:funcanforms}
    Let $\sta$ be a value environment and $\G$ be an inert, \tnew{well-formed} environment such that $\wfDft$.
    If $\typDft {\new{p}} {\tForall x T U}$ then there exists a type $T'$ and a term $t$ such that
        i) $\sta \vdash \new{p \lookupSymbol^{*}} \tLambda x {T'} t$,
        ii) $\subDft T {T'}$, and
        iii) $\typ {\extendG x T} t U$.
\end{lemma}
\noindent
This simple statement hides an intricate induction hypothesis and a long, tedious
proof, since it needs to reason precisely about function, object, and singleton types and
across all seven typing relations in the stratification of typing.

\section{Examples}\label{sec:examples}
In this section, we present three \pdot program examples that illustrate
different features of the calculus. All of the programs were formalized and
typechecked in Coq.

To make the examples easier to read, we simplify the notation for objects
$\tNew x U {d}$
by removing type
annotations where they can be easily inferred, yielding a new notation
$\nu(x\Rightarrow {d'})$:
\begin{itemize}
	\item a type definition $\set{A=T}$ can be only typed with $\tTypeDec A T T$,
		so we will skip type declarations;
	\item in a definition $\set{a=p}$, the field $a$ is assigned a path and can be only typed with 
		a singleton type; we will therefore skip the type $\tFldDec{a}{\single p}$;
	\item in a definition $\set{a=\tNew x T d}$, $a$ is assigned an object that must
		be typed with $\tRec x T$; since we can infer $T$ by looking at the object definitions,
		we will skip the typing $\tFldDec a {\tRec x T}$;
	\item we inline the type of abstractions into the field definition (e.g. $\set{a\colon\tForall x T U=\tLambda x T t}$).
\end{itemize}
For readability we will also remove the curly braces around object definitions 
and replace the $\wedge$ delimiters with semicolons.
As an example for our abbreviations, 
the object
\small
\begin{align*}
	\nu(x\colon &\tTypeDec A T T \wedge 
	            \tFldDec a {\single p}\wedge
	            \tFldDec b {\tRec y U}\wedge
	            \tFldDec c {\tForall z S V})\\
	            &\set{A=T}\wedge
	            \set{a=p}\wedge
	            \set{b=\tNew y U {d}}\wedge
	            \set{c=\tLambda{z}{S'}{t}}
\end{align*}
\normalsize
will be encoded as
$$
	\nu(x\Rightarrow\ 
		A=T;\ 
		a=p;\ 
		b=\nu(y\Rightarrow d');
		c\colon \tForall{z}{S}{V}=\tLambda{z}{S'}{t}
	)
$$
where $\nu(y\Rightarrow d')$ is the abbreviated version
of $\tNew y U {d}$.

\subsection{Class Encodings}

Fully path-dependent types allow \pdot to define encodings for Scala's
module system and classes, as we will see in the examples below.

In Scala, declaring a \code{class A(\textit{args})} automatically defines both a type \code A for
the class and a constructor for \code A with parameters \code{\textit{args}}.
We will encode such a Scala class in \pdot as a type member $A$
and a method \code{newA} that returns an object of type $A$:
\begin{multicols}{2}\noindent
	\openup-2\jot
	\begin{align*}
	\nu(p&\Rightarrow\\
	&A=\tRec {\this} {\tFldDec {\codeeq{foo}} {\forall(\_)\,U}} ;\\
	&\codeeq{newA}\colon \tForall x U {p.A}\\
	&\phantom{A}={\nu(\this)\set{\codeeq{foo}=
			\lambda(\_).\,x}})
	\end{align*}
	\begin{lstlisting}[language=Scala,basicstyle=\scalalst]
	package p {
	    class A(x: U) {
	        def foo: U = x 
	    }}
	\end{lstlisting}
\end{multicols}
\noindent
To encode subtyping we use type intersection. For example, we can 
define a class \code B that extends \code A as follows:

\begin{multicols}{2}\noindent
	\openup-2\jot
	\begin{align*}
	\nu(p&\Rightarrow\\
	&A=\tRec {\this} {\tFldDec {\codeeq{foo}} {\forall(\_)\,U}} ;\\
	&\codeeq{newA}\colon \tForall x U {p.A}\\
	&\phantom{A}={\nu(\this)\set{\codeeq{foo}=
			\lambda(\_).\,x}})
	\end{align*}
	\begin{lstlisting}[language=Scala,basicstyle=\scalalst]
	package p {
	    class A(x: U) {
	        def foo: U = x 
	    }}
	\end{lstlisting}
\end{multicols}

\subsection{Lists}

As an example to illustrate that \pdot supports the type abstractions of DOT we
formalize the covariant-list library by~\citet{wadlerfest} in \pdot, presented in
Figure~\ref{fig:examples} a).
The encoding defines \code{List} as a data type with an element type $A$
and methods \code{head} and \code{tail}.
The library contains
\code{nil} and \code{cons} fields for creating lists.
To soundly formalize the list example, we encode \code{head} and \code{tail}
as \textit{methods} (\code{def}s) as opposed to \code{val}s by wrapping them in lambda abstractions,
as discussed in Section~\ref{sec:fieldsmethods}.
This encoding also corresponds to the Scala standard library
where \code{head} and \code{tail} are \code{def}s and
not \code{val}s, and hence one cannot perform a type selection on them.

By contrast, the list example by~\citet{wadlerfest} encodes \code{head} and \code{tail}
as fields without wrapping their results in functions.
For a DOT that supports paths, such an encoding is unsound because it
violates the property that paths to objects with type members
are acyclic.
In particular, since no methods should be invoked on \code{nil}, its \code{head} and \code{tail}
methods are defined as non-terminating loops, and \code{nil}'s element type is 
instantiated to $\bot$.
If we allowed \code{nil.head} to have type $\bot$ then since $\bot<:\tTypeDec A \top \bot$,
we could derive $\top<:\code{nil.head.A}<:\bot$.
\subsection{Mutually Recursive Modules}\label{sec:mrt}
\begin{wide-rules}

\newcommand{\List}{\typeMember{List}}
\newcommand{\A}{\typeMember{A}}
\newcommand{\nil}{\termMember{nil}}
\newcommand{\cons}{\termMember{cons}}
\newcommand{\head}{\termMember{head}}
\newcommand{\tail}{\termMember{tail}}
\newcommand{\sci}{\selfVar{sci}}
\newcommand{\self}{\selfVar{self}}
\newcommand{\result}{\boundVar{result}}
\newcommand{\resultTwo}{\boundVar{result'}}
\newcommand{\hd}{\boundVar{hd}}
\newcommand{\tl}{\boundVar{tl}}

\newcommand{\phantomThirty}{\phantom{\nil}}
\newcommand{\phantomThirtyOne}{\phantomThirty
    \phantom{= \lambda(x \colon \tTypeDec{\A}{\bot}{\top})\,}}
\newcommand{\phantomThirtyTwo}{\phantom{\cons}}
\newcommand{\phantomThirtyThree}{\phantomThirty\phantom{=\lambda(x \colon }}
\newcommand{\phantomThirtyFour}{\phantomThirtyThree\phantom{\nu(\self\Rightarrow\ \ }}
\newcommand{\phantomThirtyFive}{\phantomThirtyOne\phantom{\Let\,\result=\nu(\self\Rightarrow\ \ }}
\newcommand{\phantomThirtySix}{\phantomThirtyFour\phantom{\Let\,\result=}}
{\small\textsf{a) A \textcolor{purple}{covariant list} library in \pdot}}\\
\begin{align*}
\nu({\sci}\Rightarrow\
   &\List=\hleqq{\mu(\self\colon \tTypeDec{\A}{\bot}{\top} \wedge
                          \tFldDec{\head}{\forall(\_)\,\self.\A} \wedge 
                          \tFldDec{\tail}{\forall(\_)\,
                              (\sci.\List\wedge\tTypeDec{\A}{\bot}{\self.\A})})}; \\
   &\nil
    \colon\tForall x {\tTypeDec{\A}{\bot}{\top}} {\hleqq{\sci.\List\wedge\tTypeDec{\A}{\bot}{\bot}}}
            \\ &\phantomThirty
    = \lambda(x \colon \tTypeDec{\A}{\bot}{\top})\,
      \Let\,\result=\nu(\self\Rightarrow 
        \A=\bot; \\ &\phantomThirtyFive
        \head
            \colon \tForall{y}{\top}{\self.\A}
            =\lambda(y\colon\top)\,\self.\head\,y; \\ & \phantomThirtyFive
        \tail
            \colon \tForall{y}{\top}{(\sci.\List\wedge\self.\A)}
            =\lambda(y\colon\top)\,\self.\tail\,y)\\ & \phantomThirtyOne
      \In\,\result
    ;\\
   &\cons
    \colon
        \forall(x \colon \tTypeDec{\A}{\bot}{\top})\,
        \forall(\hd\colon x.\A)\,
        \forall(\tl\colon \sci.\List\wedge \tTypeDec \A \bot {x.\A})\,
        \hleqq{\sci.\List\wedge\tTypeDec{\A}{\bot}{x.\A}}
        \\ &\phantomThirtyTwo
    = \lambda(x \colon 
        \tTypeDec{\A}{\bot}{\top})\,
      \lambda(\hd\colon x.\A)\,
      \lambda(\tl\colon \sci.\List\wedge \tTypeDec \A \bot {x.\A})\\ & \phantomThirtyThree
        \Let\,\result=
            \nu(\self\Rightarrow 
                \A=x.\A; \\ &\phantomThirtySix
                \head
                    \colon \forall(\_)\,\self.\A
                    =\lambda\_.\,\hd \\ & \phantomThirtySix
                \tail
                    \colon \forall(\_)\,(\sci.\List\wedge\self.\A)
                    =\lambda\_.\,\tl)\\ & \phantomThirtyThree
        \In\,\result)
\end{align*}
\normalsize
\newcommand{\Type}{{\typeMember{Type}}}
\newcommand{\Symbol}{\typeMember{Symbol}}
\newcommand{\ssymbol}{\termMember{newSymbol}}
\newcommand{\types}{\termMember{types}}
\newcommand{\symbols}{\termMember{symbols}}
\newcommand{\typeCstr}{\termMember{newType}}
\newcommand{\symb}{\termMember{symb}}
\newcommand{\tpe}{\termMember{tpe}}
\newcommand{\denotation}{\termMember{denotation}}
\newcommand{\srcPos}{\termMember{srcPos}}
\newcommand{\dotty}{\selfVar{dotty}}
\newcommand{\dc}{\selfVar{dc}}
\newcommand{\typess}{\selfVar{types}}
\newcommand{\symbolss}{\selfVar{symbols}}
\newcommand{\resultPath}{\boundVar{result}}
\newcommand{\resultTwoPath}{\boundVar{result'}}

\newcommand{\phantomSeven}{\phantom{\nu(\dc\Rightarrow\,}}
\newcommand{\phantomNine}{\phantomThree\phantom{\{\ssymbol\colon \forall}}
\newcommand{\phantomTen}{\phantomSeven\phantom{\types=\nu(\typess\Rightarrow\ \ }}
\newcommand{\phantomEleven}{\phantomTen\phantom{\typeCstr}}
\newcommand{\phantomTwelve}{\phantomEleven\phantom{\colon \forall}}
\newcommand{\phantomThirteen}{\phantomEleven\phantom{=\lambda(s\colon\,}}
\newcommand{\phantomFourteen}{\phantomThirteen\phantom{\code{let}\,}}
\newcommand{\phantomFifteen}{\phantomFourteen\phantom{\code{result'}=\nu(r\Rightarrow\ \ }}
\newcommand{\phantomSixteen}{\phantomSeven\phantom{\symbols=\nu(\symbolss\Rightarrow\ }}
\newcommand{\phantomSeventeen}{\phantomSixteen\phantom{\ssymbol}}
\newcommand{\phantomEighteen}{\phantomSeventeen\phantom{\colon \forall}}
\newcommand{\phantomNineteen}{\phantomSeventeen\phantom{=\lambda(t\colon }}
\newcommand{\phantomTwenty}{\phantomNineteen\phantom{\code{let}\ }}
\newcommand{\phantomTwentyone}{\phantomTwenty\phantom{\code{result'}=\nu(r\Rightarrow\ \,}}
{\small\textsf{b) Mutually recursive types in a compiler package: \textcolor{blue}{fully path-dependent types}}}\\
\begin{align*}
&
\nu(\dc\Rightarrow
    \types=\nu(\typess\Rightarrow
        \Type=\tRec {\this} {\tFldDec{\symb}{\hleq{\dc.\symbols.\Symbol}}}; \\ &\phantomTen
        \typeCstr
            \colon \forall
                (s \colon \hleq{\dc.\symbols.\Symbol})\,\typess.\Type\\ &\phantomEleven
            =\lambda(s\colon 
                \hleq{\dc.\symbols.\Symbol})\\ &\phantomThirteen
                    \Let\ 
                        \resultTwoPath=\nu(\this\Rightarrow 
                              \symb
                                =s) \ \In\ \resultTwoPath
         );\\&\phantomSeven
    \symbols=\nu(\symbolss\Rightarrow
        \Symbol=\tRec {\this} {\tFldDec{\tpe}{\hleq{\dc.\types.\Type}}}; \\ &\phantomSixteen
        \ssymbol
            \colon \forall
                (t \colon \hleq{\dc.\types.\Type})\,\symbolss.\Symbol \\ &\phantomSeventeen
            =\lambda(t\colon 
                \hleq{\dc.\types.\Type})\\ &\phantomNineteen
                    \Let\ 
                        \resultTwoPath=\nu(\this\Rightarrow 
                            \tpe
                                =t)
                                \ \In\ \resultTwoPath
    ))
\end{align*}
\newcommand{\incr}{{\termMember{incr}}}
\newcommand{\decr}{{\termMember{decr}}}
\newcommand{\newD}{{\termMember{newD}}}

\newcommand{\phantomTwentythree}{\phantom{\Let\ \codeeq{pkg}=\nu(p\Rightarrow\ \ }}
\newcommand{\phantomTwentytwo}{\phantomTwentythree\phantom{d \colon \forall(\_)}}
{\small\textsf{c) \textcolor{purple}{Chaining} method calls using \textcolor{purple}{singleton types}}}\\
\small
\begin{align*}
	&
	 \Let\ \codeeq{pkg}=\nu(p\Rightarrow\ 
		C =\tRec \this {\tFldDec \incr {\hleqq{\single \this}}};\\&\phantomTwentythree
		D =\tRec \this {p.C \wedge \tFldDec \decr {\hleqq{\single \this}}};\\&\phantomTwentythree
		\newD \colon \forall(\_)\ 
			{p.D}=\lambda\,\_.\\ &\phantomTwentytwo
		    \Let\ 
		    	\codeeq{result}=\nu(\this\Rightarrow 
					\incr = \this;
					\decr = \this)\\ &\phantomTwentytwo
		    \In\ \codeeq{result})\\
    &\In\ \Let\ d=\codeeq{pkg}.\newD\ \_\\
    &\In\ \hleqq{d.\incr.\decr}\ \ 
\end{align*}
\normalsize
\caption{Example \pdot encodings}\label{fig:examples}
\end{wide-rules}
The second example, presented in Figure~\ref{fig:examples} b),
illustrates \pdot's ability to use path-dependent types
of arbitrary length. It formalizes the compiler example from Section~\ref{sec:intro}
in which the nested classes \code{Type} and \code{Symbol} recursively reference each other.
\subsection{Chaining methods with singleton types}

The last example focuses on \pdot's ability to use singleton types as they
are motivated by~\citet{cake}.
An example from that paper
introduces a class $C$ with an \code{incr} method that increments a mutable
integer field and returns the object itself (\code{this}). A class $D$ extends $C$ and
defines an analogous \code{decr} method. By declaring the return types of
the \code{incr} and \code{decr} methods as singleton types,
we can invoke a chain of method calls
$d.\code{incr}.\code{decr}$ on an object $d$ of type $D$. If \code{incr}
were declared with type $C$ this would not have been possible since $C$ does not
have a \code{decr} member.

Our formalization of the example is displayed in Figure~\ref{fig:examples} c).
The original example contains a class $A$ that declares a mutable integer field $x$ and a 
method \code{incr} that increases $x$ and returns the object itself. A class $B$ extends $A$ 
with an analogous \code{decr} method that decreases $x$. The example shows how we can invoke 
a chain of \code{incr} and \code{decr} methods on an object of type $B$ using singleton 
types: if $A.\code{incr}$ returned an object of type $A$ this would be impossible, so the method's 
return type is \code{this.type}. Since pDOT does not support mutation, our example excludes the 
mutation side effect of the original example which is there to make the example more 
practical.
\section{Related Work}\label{sec:related}

This section reviews work related to formalizing Scala with support for fully path-dependent types.

\subsection{Early Class-based Scala Formalizations}


\newcommand{\fs}{FS$_{alg}$\xspace}

Several predecessors of the DOT calculus support path-dependent types on paths of arbitrary length.
The first Scala formalization, $\nu$Obj~\cite{nuobj}, is a nominal, class-based calculus
with a rich set of language features that formalizes object-dependent type members.
Two subsequent calculi, Featherweight Scala (\fs)~\cite{fs} and Scalina~\cite{scalina},
build on $\nu$Obj to establish Scala formalizations with algorithmic typing and
with full support for higher-kinded types.
All three calculi support paths of arbitrary length, singleton types, and abstract type members.
Whereas \fs supports type-member selection directly on paths,
$\nu$Obj and Scalina allow selection $T\#A$ on types.
A path-dependent type $p.A$ can thus be encoded as a selection on a singleton type: $\single p\#A$.
$\nu$Obj is the only of the above calculi that comes with a type-safety proof. The proof is non-mechanized.

Both \pdot and these calculi prevent type selections on non-terminating paths.
$\nu$Obj achieves this through a \textit{contraction} requirement that prevents a term on the right-hand side of a
definition from referring to the definition's self variable. At the same time, recursive calls can be encoded in $\nu$Obj by
referring to the self variable from a nested class definition.
\fs ensures that paths are normalizing through a cycle detection mechanism that
ensures that a field selection can appear only once as part of a path.
Scalina avoids type selection $T\#A$ on a non-terminating type $T$ by explicitly requiring $T$ to be of
a \textit{concrete} kind, which means that $T$ expands to a structural type $R$ that contains a type member $A$.
Although Scalina allows $A$ to have upper and lower bounds, bad bounds are avoided because $A$ also needs to
be immediately initialized with a type $U$ that conforms to $A$'s bounds, which is more restrictive than DOT.
In \pdot, it is possible to create cyclic paths but impossible to do a type selection on them
because as explained in Section~\ref{sec:typerestrict}, cyclic paths that can appear in a concrete execution
context cannot be typed with a type-member declaration.

A difference between \pdot and the above calculi is that to ensure type soundness, paths in \pdot are
normal form. This is necessary to ensure that each object has a name, as explained in Section~\ref{sec:pathsnormalform}.
$\nu$Obj and \fs achieve type safety in spite of reducing paths by allowing field selection only on variables.
This way, field selections always occur on named objects.
Scalina does not require objects to be tied to names. In particular, its field selection rule \rn{E\_Sel}
allows a field selection $\code{new}\,T.a$ on an object if $T$ contains a field definition $\set{a=s}$.
The selection reduces to $\tSubst {\this} {\code{new}\,T} s$,
i.e. each occurrence of the self variable is replaced with a copy of $\code{new}\,T$.

A second difference to \pdot is the handling of singleton types.
In order to reason about a singleton type $\single p$, $\nu$Obj, \fs, and Scalina use several recursively
defined judgments (membership, expansion, and others) that rely on analyzing the shape and well-formedness
of the type that $p$ expands to.
By contrast, \pdot contains one simple \rn{Sngl-Trans} rule that allows a path to inherit the
type of its alias.
On the other hand, \pdot has the shortcoming that singleton typing is not reflexive.
Unlike in the above systems and in Scala, \pdot lacks a type axiom $\typDft p {\single p}$.
Such a rule would undermine the anti-symmetry of path aliasing which is essential to the safety proof.

None of the other calculi have to confront the problem of bad bounds.
Unlike DOT, \pdot, and Scala, $\nu$Obj and \fs
do not support lower bounds of type members and have no unique upper and lower bounds on types.
Scalina does have top and bottom types and supports bounds through interval kinds, but
it avoids bad bounds by requiring types on which selection occurs to be concrete.
In addition, it is unknown whether Scalina and \fs are sound.

Finally, the three type systems are nominal and class based, and include a large set of language features that are present in Scala.
DOT is a simpler and smaller calculus that abstracts over many of the design
decisions of the above calculi. Since DOT aims to be a base for experimentation
with new language features,
it serves well as a minimal core calculus for languages with type members,
and the goal of \pdot is to generalize DOT to fully path-dependent types.

\subsection{DOT-like Calculi}

\citet{fool12} present the first version of a DOT calculus. It includes type intersection, recursive types,
unique top and bottom types, type members with upper and lower bounds, and
path-dependent types on paths of arbitrary length.
This version of DOT has explicit support for fields (\code{val}s) and methods (\code{def}s).
Fields must be initialized to variables, which prevents the creation of non-terminating paths
(since that would require initializing fields to paths), but it also limits expressivity.
Specifically, just like in DOT by~\citet{wadlerfest}, path-dependent types cannot refer to nested modules
because modules have to be created through methods, and method invocations cannot be part of
a path-dependent type.
The calculus is not type-safe, as illustrated by multiple counterexamples in the paper. In particular,
this version of DOT does not track path equality which, as explained in the paper, breaks preservation.

To be type-safe, DOT must ensure that path-dependent types are invoked only on terminating paths. A possible strategy to ensure a sound DOT with support for paths
is to investigate the conditions under which terms terminate, and
to impose these conditions on the paths that participate in type selections.
To address these questions, \citet{normalization} present a Coq-mechanized semantic proof of strong normalization for the \dsub calculus.
\dsub is a generalization of System \fsub with lower- and upper-bounded type tags and variable-dependent types.
The paper shows that recursive objects constitute the feature that enables recursion and hence
Turing-completeness in DOT.
Since \dsub lacks recursive objects, it
is strongly normalizing.
Furthermore, the lack of objects and fields implies that this version of \dsub can only express paths that are variables.

\citet{patheq} present $\pi$DOT, a strongly normalizing version of a \dsub 
without top and bottom types but with support for paths of arbitrary length.
$\pi$DOT keeps track of path aliasing through path-equivalence sets, and the paper also mentions the possibility of using singleton
types to formalize path equality.
Like the calculus by~\citet{normalization},
this version of \dsub is strongly normalizing due to the lack of recursive self variables.
This guarantees that paths are acyclic. It also ensures that due to the lack
of recursion elimination, reducing paths preserves soundness
(unlike in \pdot, as explained in Section~\ref{sec:pathsnormalform}).
$\pi$DOT comes with a non-mechanized soundness proof.

By contrast with these two papers, our work proposes a Turing-complete generalization with paths of arbitrary length
of the full DOT calculus, which includes recursive objects and type intersections.

\subsection{Other Related Languages and Calculi}

Scala's module system shares many similarities with languages of the ML family.
Earlier presentations of ML module systems~\cite{mlharper,mlleroy,mldreyer}
allow fine-grained control over type abstractions and code
reuse but do not support mutually recursive modules, and
separate the language of terms from the language of modules.
\emph{MixML} extends the essential features of these type systems
with the ability to do both \emph{hierarchical} and \emph{mixin} composition
of modules~\cite{mixml}.
The language supports \emph{recursive} modules which can be packaged
as \emph{first-class values}.
The expressive power of \emph{MixML}'s module system, plus support for
decidable type-checking
requires a set of constraints
on the \emph{linking} (module mixin) operation that restrict recursion between modules,
including a total order on \emph{label paths},
and
yields a complex
type system that closely models actual implementations of ML.

\citet{fing} and \citet{fingfirst2} address the inherent complexity of ML module systems
by presenting encodings of an ML-style module language
into System F$_{\omega}$. The latter paper presents \emph{1ML}, a
concise version of ML that fully unifies the language of modules with the language of terms.
However, both formalizations exclude recursive modules.

A type system that distinguishes types based on the runtime values of
their enclosing objects was first introduced by \citet{family}
in the context of \emph{family polymorphism}.
Notably, family polymorphism is supported by \emph{virtual classes},
which can be inherited and overriden within different objects
and whose concrete implementation is resolved at runtime.
Virtual classes are
supported in the \emph{Beta} and \emph{gbeta} programming
languages~\cite{beta,gbeta} (but not in Scala in which classes are
statically resolved at compile time) and
formalized by the \emph{vc} and \emph{Tribe} calculi~\cite{vc,tribe}.
Paths in \emph{vc}
are relative to \code{this} and consist of a
sequence of \code{out} keywords, which refer to enclosing objects,
and field names. 
To track path equality, \emph{vc} uses a
normalization function that converts paths to a canonical representation,
and to rule out cyclic paths it defines a partial order
on declared names.
\emph{Tribe}'s paths can be both relative or absolute: they can start with a variable, and 
they can intermix class and object references.
The calculus uses singleton types to track path equality and rules out cyclic paths by disallowing
cyclic dependencies in its inheritance relation.

A difference between \pdot and all of \emph{vc}, \emph{Tribe}, and
the ML formalizations is that \pdot does not impose any orderings
on paths, and fully supports recursive references
between objects and path-dependent types.
In addition, \pdot's ability to define type members with both lower
and upper bounds introduces a complex source of unsoundness in the form
of \emph{bad bounds}
(alas, the cost for its expressiveness is 
that \pdot's type system is likely not decidable, as discussed below).
Yet, by being mostly structurally typed,
without having to model initialization and inheritance,
\pdot remains general and small.
Finally, by contrast to the above, \pdot has comes with a mechanized type-safety proof.

\subsection{Decidability}
The baseline DOT calculus to which we add the path extensions is widely
conjectured to have undecidable typechecking because it includes the
features of \fsub, for which typechecking is undecidable~\cite{decidability}. 
\citet{oopsla16} give a mapping from \fsub to \dsub, a simpler calculus than DOT, 
and prove that if the \fsub term is typeable then so is the \dsub term, but the only-if 
direction and therefore the decidability of \dsub and DOT remain open problems, 
subject of active research. The open question of decidability of DOT needs to be 
resolved before we can consider decidability of pDOT.

We believe that \pdot does not introduce additional sources of undecidability
into DOT. One feature of \pdot that might call this into question is
singleton types.
In particular, \citet{stoneharper} study systems of singleton kinds that reason about 
types with non-trivial reduction rules, yet it remains decidable which 
types reduce to the same normal form. The singleton types of both Scala and 
\pdot are much simpler and less expressive in that only assignment of an object 
between variables and paths is allowed, but the objects are not arbitrary terms and do 
not reduce. Thus, the Scala and \pdot singleton types only need to track sequences of 
assignments. Thus, although decidability of \pdot is unknown because it is unknown 
for DOT, the singleton types that we add in \pdot are unlikely to affect decidability 
because they are significantly less expressive than the singleton types studied by 
Stone and Harper.

\section{Conclusion}\label{sec:conclusion}

The DOT calculus was designed as a small core calculus to model Scala's type system with a focus on path-dependent types.
However, DOT can only model types that depend on variables, which
significantly under-approximates the behaviour of Scala programs.
Scala and, more generally, languages with type members need to rely on fully path-dependent types to encode
the possible type dependencies in their module systems without restrictions.
Until now, it was unclear whether combining the fundamental features of languages with path-dependent types, namely
bounded abstract type members, intersections, recursive objects, and paths of arbitrary length is type-safe.

This paper proposes \pdot, a calculus that generalizes DOT with support for paths of arbitrary length.
The main insights of \pdot are to represent object identity through paths,
to ensure that well-typed acyclic paths always represent values,
to track path equality with singleton types,
and to eliminate type selections on cyclic paths through precise object typing.
\pdot allows us to use the full potential of path-dependent types. 
\pdot comes with a type-safety proof and motivating examples for fully path-dependent types and singleton types
that are mechanized in Coq.

\begin{acks}
We would like to thank Martin Odersky for suggesting the idea of extending DOT
with paths of arbitrary length, and for the helpful discussions on early variants of pDOT.
We thank Jaemin Hong, Abel Nieto, and the anonymous reviewers
for their careful reading of our paper and the insightful suggestions that greatly helped improve it.
We thank Lu Wang and Yaoyu Zhao for their contributions to the Coq proof.
We thank Paolo Giarrusso and Ifaz Kabir for their thoughtful proofreading, and the helpful
discussions on the DOT calculus which improved our understanding of the the expressiveness
and limitations of DOT.
We had other helpful discussions on DOT with Zhong Sheng, Sukyoung Ryu, Derek Dreyer, Ilya Sergey, and Prabhakar Ragde.
This research was supported by the
Natural Sciences and Engineering Research Council of Canada.
\end{acks}


\begin{thebibliography}{31}


\ifx \showCODEN    \undefined \def \showCODEN     #1{\unskip}     \fi
\ifx \showDOI      \undefined \def \showDOI       #1{#1}\fi
\ifx \showISBNx    \undefined \def \showISBNx     #1{\unskip}     \fi
\ifx \showISBNxiii \undefined \def \showISBNxiii  #1{\unskip}     \fi
\ifx \showISSN     \undefined \def \showISSN      #1{\unskip}     \fi
\ifx \showLCCN     \undefined \def \showLCCN      #1{\unskip}     \fi
\ifx \shownote     \undefined \def \shownote      #1{#1}          \fi
\ifx \showarticletitle \undefined \def \showarticletitle #1{#1}   \fi
\ifx \showURL      \undefined \def \showURL       {\relax}        \fi
\providecommand\bibfield[2]{#2}
\providecommand\bibinfo[2]{#2}
\providecommand\natexlab[1]{#1}
\providecommand\showeprint[2][]{arXiv:#2}

\bibitem[\protect\citeauthoryear{Amin, Gr{\"{u}}tter, Odersky, Rompf, and
  Stucki}{Amin et~al\mbox{.}}{2016}]%
        {wadlerfest}
\bibfield{author}{\bibinfo{person}{Nada Amin}, \bibinfo{person}{Samuel
  Gr{\"{u}}tter}, \bibinfo{person}{Martin Odersky}, \bibinfo{person}{Tiark
  Rompf}, {and} \bibinfo{person}{Sandro Stucki}.}
  \bibinfo{year}{2016}\natexlab{}.
\newblock \showarticletitle{The Essence of Dependent Object Types}. In
  \bibinfo{booktitle}{\emph{A List of Successes That Can Change the World -
  Essays Dedicated to Philip Wadler on the Occasion of His 60th Birthday}}.
  \bibinfo{pages}{249--272}.
\newblock


\bibitem[\protect\citeauthoryear{Amin, Moors, and Odersky}{Amin
  et~al\mbox{.}}{2012}]%
        {fool12}
\bibfield{author}{\bibinfo{person}{Nada Amin}, \bibinfo{person}{Adriaan Moors},
  {and} \bibinfo{person}{Martin Odersky}.} \bibinfo{year}{2012}\natexlab{}.
\newblock \showarticletitle{Dependent {O}bject {T}ypes}. In
  \bibinfo{booktitle}{\emph{International Workshop on Foundations of
  Object-Oriented Languages (FOOL 2012)}}.
\newblock


\bibitem[\protect\citeauthoryear{Amin and Rompf}{Amin and Rompf}{2017}]%
        {popl17}
\bibfield{author}{\bibinfo{person}{Nada Amin} {and} \bibinfo{person}{Tiark
  Rompf}.} \bibinfo{year}{2017}\natexlab{}.
\newblock \showarticletitle{Type soundness proofs with definitional
  interpreters}. In \bibinfo{booktitle}{\emph{Proceedings of the 44th {ACM}
  {SIGPLAN} Symposium on Principles of Programming Languages, {POPL} 2017,
  Paris, France, January 18-20, 2017}}. \bibinfo{pages}{666--679}.
\newblock


\bibitem[\protect\citeauthoryear{Amin, Rompf, and Odersky}{Amin
  et~al\mbox{.}}{2014}]%
        {oopsla14}
\bibfield{author}{\bibinfo{person}{Nada Amin}, \bibinfo{person}{Tiark Rompf},
  {and} \bibinfo{person}{Martin Odersky}.} \bibinfo{year}{2014}\natexlab{}.
\newblock \showarticletitle{Foundations of path-dependent types}. In
  \bibinfo{booktitle}{\emph{Proceedings of the 2014 {ACM} International
  Conference on Object Oriented Programming Systems Languages {\&}
  Applications, {OOPSLA} 2014, part of {SPLASH} 2014, Portland, OR, USA,
  October 20-24, 2014}}. \bibinfo{pages}{233--249}.
\newblock


\bibitem[\protect\citeauthoryear{Amin and Tate}{Amin and Tate}{2016}]%
        {null}
\bibfield{author}{\bibinfo{person}{Nada Amin} {and} \bibinfo{person}{Ross
  Tate}.} \bibinfo{year}{2016}\natexlab{}.
\newblock \showarticletitle{Java and {S}cala's type systems are unsound: the
  existential crisis of null pointers}. In
  \bibinfo{booktitle}{\emph{Proceedings of the 2016 {ACM} {SIGPLAN}
  International Conference on Object-Oriented Programming, Systems, Languages,
  and Applications, {OOPSLA} 2016, part of {SPLASH} 2016, Amsterdam, The
  Netherlands, October 30 - November 4, 2016}}. \bibinfo{pages}{838--848}.
\newblock


\bibitem[\protect\citeauthoryear{Aydemir, Chargu{\'{e}}raud, Pierce, Pollack,
  and Weirich}{Aydemir et~al\mbox{.}}{2008}]%
        {ln}
\bibfield{author}{\bibinfo{person}{Brian~E. Aydemir}, \bibinfo{person}{Arthur
  Chargu{\'{e}}raud}, \bibinfo{person}{Benjamin~C. Pierce},
  \bibinfo{person}{Randy Pollack}, {and} \bibinfo{person}{Stephanie Weirich}.}
  \bibinfo{year}{2008}\natexlab{}.
\newblock \showarticletitle{Engineering formal metatheory}. In
  \bibinfo{booktitle}{\emph{Proceedings of the 35th {ACM} {SIGPLAN-SIGACT}
  Symposium on Principles of Programming Languages, {POPL} 2008, San Francisco,
  California, USA, January 7-12, 2008}}. \bibinfo{pages}{3--15}.
\newblock


\bibitem[\protect\citeauthoryear{Bruce, Odersky, and Wadler}{Bruce
  et~al\mbox{.}}{1998}]%
        {virtual}
\bibfield{author}{\bibinfo{person}{Kim~B. Bruce}, \bibinfo{person}{Martin
  Odersky}, {and} \bibinfo{person}{Philip Wadler}.}
  \bibinfo{year}{1998}\natexlab{}.
\newblock \showarticletitle{A Statically Safe Alternative to Virtual Types}. In
  \bibinfo{booktitle}{\emph{ECOOP'98 - Object-Oriented Programming, 12th
  European Conference}}. \bibinfo{pages}{523--549}.
\newblock


\bibitem[\protect\citeauthoryear{Clarke, Drossopoulou, Noble, and
  Wrigstad}{Clarke et~al\mbox{.}}{2007}]%
        {tribe}
\bibfield{author}{\bibinfo{person}{Dave Clarke}, \bibinfo{person}{Sophia
  Drossopoulou}, \bibinfo{person}{James Noble}, {and} \bibinfo{person}{Tobias
  Wrigstad}.} \bibinfo{year}{2007}\natexlab{}.
\newblock \showarticletitle{Tribe: a simple virtual class calculus}. In
  \bibinfo{booktitle}{\emph{Proceedings of the 6th International Conference on
  Aspect-Oriented Software Development, {AOSD} 2007, Vancouver, British
  Columbia, Canada, March 12-16, 2007}}. \bibinfo{pages}{121--134}.
\newblock


\bibitem[\protect\citeauthoryear{Cremet, Garillot, Lenglet, and Odersky}{Cremet
  et~al\mbox{.}}{2006}]%
        {fs}
\bibfield{author}{\bibinfo{person}{Vincent Cremet},
  \bibinfo{person}{Fran{\c{c}}ois Garillot}, \bibinfo{person}{Sergue{\"{\i}}
  Lenglet}, {and} \bibinfo{person}{Martin Odersky}.}
  \bibinfo{year}{2006}\natexlab{}.
\newblock \showarticletitle{A Core Calculus for {Scala} Type Checking}. In
  \bibinfo{booktitle}{\emph{Mathematical Foundations of Computer Science, 31st
  International Symposium, Slovakia}}.
\newblock


\bibitem[\protect\citeauthoryear{Documentation}{Documentation}{2018}]%
        {stable}
\bibfield{author}{\bibinfo{person}{Scala Documentation}.}
  \bibinfo{year}{2018}\natexlab{}.
\newblock \bibinfo{title}{Paths}.
\newblock
\newblock
\urldef\tempurl%
\url{https://www.scala-lang.org/files/archive/spec/2.11/03-types.html#paths}
\showURL{%
Retrieved February 26, 2019 from \tempurl}


\bibitem[\protect\citeauthoryear{Dreyer, Crary, and Harper}{Dreyer
  et~al\mbox{.}}{2003}]%
        {mldreyer}
\bibfield{author}{\bibinfo{person}{Derek Dreyer}, \bibinfo{person}{Karl Crary},
  {and} \bibinfo{person}{Robert Harper}.} \bibinfo{year}{2003}\natexlab{}.
\newblock \showarticletitle{A type system for higher-order modules}. In
  \bibinfo{booktitle}{\emph{Conference Record of {POPL} 2003: The 30th
  {SIGPLAN-SIGACT} Symposium on Principles of Programming Languages, New
  Orleans, Louisisana, USA, January 15-17, 2003}}. \bibinfo{pages}{236--249}.
\newblock


\bibitem[\protect\citeauthoryear{Ernst}{Ernst}{1999}]%
        {gbeta}
\bibfield{author}{\bibinfo{person}{Erik Ernst}.}
  \bibinfo{year}{1999}\natexlab{}.
\newblock \emph{\bibinfo{title}{gbeta -- a Language with Virtual Attributes,
  Block Structure, and Propagating, Dynamic Inheritance}}.
\newblock \bibinfo{thesistype}{Ph.D. Dissertation}. \bibinfo{school}{Department
  of Computer Science, University of Aarhus, \AA{}rhus, Denmark}.
\newblock


\bibitem[\protect\citeauthoryear{Ernst}{Ernst}{2001}]%
        {family}
\bibfield{author}{\bibinfo{person}{Erik Ernst}.}
  \bibinfo{year}{2001}\natexlab{}.
\newblock \showarticletitle{Family Polymorphism}. In
  \bibinfo{booktitle}{\emph{{ECOOP} 2001 - Object-Oriented Programming, 15th
  European Conference, Budapest, Hungary, June 18-22, 2001, Proceedings}}.
  \bibinfo{pages}{303--326}.
\newblock


\bibitem[\protect\citeauthoryear{Ernst, Ostermann, and Cook}{Ernst
  et~al\mbox{.}}{2006}]%
        {vc}
\bibfield{author}{\bibinfo{person}{Erik Ernst}, \bibinfo{person}{Klaus
  Ostermann}, {and} \bibinfo{person}{William~R. Cook}.}
  \bibinfo{year}{2006}\natexlab{}.
\newblock \showarticletitle{A virtual class calculus}. In
  \bibinfo{booktitle}{\emph{Proceedings of the 33rd {ACM} {SIGPLAN-SIGACT}
  Symposium on Principles of Programming Languages, {POPL} 2006, Charleston,
  South Carolina, USA, January 11-13, 2006}}. \bibinfo{pages}{270--282}.
\newblock


\bibitem[\protect\citeauthoryear{Harper and Lillibridge}{Harper and
  Lillibridge}{1994}]%
        {mlharper}
\bibfield{author}{\bibinfo{person}{Robert Harper} {and} \bibinfo{person}{Mark
  Lillibridge}.} \bibinfo{year}{1994}\natexlab{}.
\newblock \showarticletitle{A Type-Theoretic Approach to Higher-Order Modules
  with Sharing}. In \bibinfo{booktitle}{\emph{Conference Record of POPL'94:
  21st {ACM} {SIGPLAN-SIGACT} Symposium on Principles of Programming Languages,
  Portland, Oregon, USA, January 17-21, 1994}}. \bibinfo{pages}{123--137}.
\newblock


\bibitem[\protect\citeauthoryear{Hong, Park, and Ryu}{Hong
  et~al\mbox{.}}{2018}]%
        {patheq}
\bibfield{author}{\bibinfo{person}{Jaemin Hong}, \bibinfo{person}{Jihyeok
  Park}, {and} \bibinfo{person}{Sukyoung Ryu}.}
  \bibinfo{year}{2018}\natexlab{}.
\newblock \showarticletitle{Path Dependent Types with Path-equality}. In
  \bibinfo{booktitle}{\emph{Proceedings of the 9th ACM SIGPLAN International
  Symposium on Scala}} \emph{(\bibinfo{series}{Scala 2018})}.
  \bibinfo{publisher}{ACM}, \bibinfo{pages}{35--39}.
\newblock
\showISBNx{978-1-4503-5836-1}


\bibitem[\protect\citeauthoryear{Kabir and Lhot{\'a}k}{Kabir and
  Lhot{\'a}k}{2018}]%
        {constructors}
\bibfield{author}{\bibinfo{person}{Ifaz Kabir} {and}
  \bibinfo{person}{Ond{\v{r}}ej Lhot{\'a}k}.} \bibinfo{year}{2018}\natexlab{}.
\newblock \showarticletitle{$\kappa$DOT: scaling DOT with mutation and
  constructors}. In \bibinfo{booktitle}{\emph{Proceedings of the 9th ACM
  SIGPLAN International Symposium on Scala}}. ACM, \bibinfo{pages}{40--50}.
\newblock


\bibitem[\protect\citeauthoryear{Leroy}{Leroy}{1994}]%
        {mlleroy}
\bibfield{author}{\bibinfo{person}{Xavier Leroy}.}
  \bibinfo{year}{1994}\natexlab{}.
\newblock \showarticletitle{Manifest Types, Modules, and Separate Compilation}.
  In \bibinfo{booktitle}{\emph{Conference Record of POPL'94: 21st {ACM}
  {SIGPLAN-SIGACT} Symposium on Principles of Programming Languages, Portland,
  Oregon, USA, January 17-21, 1994}}. \bibinfo{pages}{109--122}.
\newblock


\bibitem[\protect\citeauthoryear{Madsen and M{\o}ller{-}Pedersen}{Madsen and
  M{\o}ller{-}Pedersen}{1989}]%
        {beta}
\bibfield{author}{\bibinfo{person}{Ole~Lehrmann Madsen} {and}
  \bibinfo{person}{Birger M{\o}ller{-}Pedersen}.}
  \bibinfo{year}{1989}\natexlab{}.
\newblock \showarticletitle{Virtual Classes: {A} Powerful Mechanism in
  Object-Oriented Programming}. In \bibinfo{booktitle}{\emph{Conference on
  Object-Oriented Programming: Systems, Languages, and Applications
  (OOPSLA'89), New Orleans, Louisiana, USA, October 1-6, 1989, Proceedings.}}
  \bibinfo{pages}{397--406}.
\newblock


\bibitem[\protect\citeauthoryear{Moors, Piessens, and Odersky}{Moors
  et~al\mbox{.}}{2008}]%
        {scalina}
\bibfield{author}{\bibinfo{person}{Adriaan Moors}, \bibinfo{person}{Frank
  Piessens}, {and} \bibinfo{person}{Martin Odersky}.}
  \bibinfo{year}{2008}\natexlab{}.
\newblock \showarticletitle{Safe type-level abstraction in Scala}. In
  \bibinfo{booktitle}{\emph{International Workshop on Foundations of
  Object-Oriented Languages (FOOL 2008)}}.
\newblock


\bibitem[\protect\citeauthoryear{Odersky, Cremet, R{\"{o}}ckl, and
  Zenger}{Odersky et~al\mbox{.}}{2003}]%
        {nuobj}
\bibfield{author}{\bibinfo{person}{Martin Odersky}, \bibinfo{person}{Vincent
  Cremet}, \bibinfo{person}{Christine R{\"{o}}ckl}, {and}
  \bibinfo{person}{Matthias Zenger}.} \bibinfo{year}{2003}\natexlab{}.
\newblock \showarticletitle{A Nominal Theory of Objects with Dependent Types}.
  In \bibinfo{booktitle}{\emph{{ECOOP} 2003 - Object-Oriented Programming, 17th
  European Conference, Darmstadt, Germany, July 21-25, 2003, Proceedings}}.
  \bibinfo{pages}{201--224}.
\newblock


\bibitem[\protect\citeauthoryear{Odersky and Zenger}{Odersky and
  Zenger}{2005}]%
        {cake}
\bibfield{author}{\bibinfo{person}{Martin Odersky} {and}
  \bibinfo{person}{Matthias Zenger}.} \bibinfo{year}{2005}\natexlab{}.
\newblock \showarticletitle{Scalable component abstractions}. In
  \bibinfo{booktitle}{\emph{Proceedings of the 20th Annual {ACM} {SIGPLAN}
  Conference on Object-Oriented Programming, Systems, Languages, and
  Applications, {OOPSLA} 2005, October 16-20, 2005, San Diego, CA, {USA}}}.
  \bibinfo{pages}{41--57}.
\newblock


\bibitem[\protect\citeauthoryear{Pierce}{Pierce}{1992}]%
        {decidability}
\bibfield{author}{\bibinfo{person}{Benjamin~C. Pierce}.}
  \bibinfo{year}{1992}\natexlab{}.
\newblock \showarticletitle{Bounded Quantification is Undecidable}. In
  \bibinfo{booktitle}{\emph{Conference Record of the Nineteenth Annual {ACM}
  {SIGPLAN-SIGACT} Symposium on Principles of Programming Languages,
  Albuquerque, New Mexico, USA, January 19-22, 1992}}.
  \bibinfo{pages}{305--315}.
\newblock


\bibitem[\protect\citeauthoryear{Rapoport, Kabir, He, and
  Lhot{\'{a}}k}{Rapoport et~al\mbox{.}}{2017}]%
        {simpl}
\bibfield{author}{\bibinfo{person}{Marianna Rapoport}, \bibinfo{person}{Ifaz
  Kabir}, \bibinfo{person}{Paul He}, {and} \bibinfo{person}{{Ond\v rej}
  Lhot{\'{a}}k}.} \bibinfo{year}{2017}\natexlab{}.
\newblock \showarticletitle{A simple soundness proof for dependent object
  types}.
\newblock \bibinfo{journal}{\emph{{PACMPL}}} \bibinfo{volume}{1},
  \bibinfo{number}{{OOPSLA}} (\bibinfo{year}{2017}),
  \bibinfo{pages}{46:1--46:27}.
\newblock


\bibitem[\protect\citeauthoryear{Rompf and Amin}{Rompf and Amin}{2016}]%
        {oopsla16}
\bibfield{author}{\bibinfo{person}{Tiark Rompf} {and} \bibinfo{person}{Nada
  Amin}.} \bibinfo{year}{2016}\natexlab{}.
\newblock \showarticletitle{Type soundness for dependent object types {(DOT)}}.
  In \bibinfo{booktitle}{\emph{Proceedings of the 2016 {ACM} {SIGPLAN}
  International Conference on Object-Oriented Programming, Systems, Languages,
  and Applications, {OOPSLA} 2016, part of {SPLASH} 2016, Amsterdam, The
  Netherlands, October 30 - November 4, 2016}}. \bibinfo{pages}{624--641}.
\newblock


\bibitem[\protect\citeauthoryear{Rossberg}{Rossberg}{2018}]%
        {fingfirst2}
\bibfield{author}{\bibinfo{person}{Andreas Rossberg}.}
  \bibinfo{year}{2018}\natexlab{}.
\newblock \showarticletitle{1ML --- Core and modules united}.
\newblock \bibinfo{journal}{\emph{Journal of Functional Programming}}
  \bibinfo{volume}{28} (\bibinfo{year}{2018}), \bibinfo{pages}{e22}.
\newblock


\bibitem[\protect\citeauthoryear{Rossberg and Dreyer}{Rossberg and
  Dreyer}{2013}]%
        {mixml}
\bibfield{author}{\bibinfo{person}{Andreas Rossberg} {and}
  \bibinfo{person}{Derek Dreyer}.} \bibinfo{year}{2013}\natexlab{}.
\newblock \showarticletitle{Mixin' Up the {ML} Module System}.
\newblock \bibinfo{journal}{\emph{{ACM} Trans. Program. Lang. Syst.}}
  \bibinfo{volume}{35}, \bibinfo{number}{1} (\bibinfo{year}{2013}),
  \bibinfo{pages}{2:1--2:84}.
\newblock


\bibitem[\protect\citeauthoryear{Rossberg, Russo, and Dreyer}{Rossberg
  et~al\mbox{.}}{2014}]%
        {fing}
\bibfield{author}{\bibinfo{person}{Andreas Rossberg},
  \bibinfo{person}{Claudio~V. Russo}, {and} \bibinfo{person}{Derek Dreyer}.}
  \bibinfo{year}{2014}\natexlab{}.
\newblock \showarticletitle{F-ing modules}.
\newblock \bibinfo{journal}{\emph{J. Funct. Program.}} \bibinfo{volume}{24},
  \bibinfo{number}{5} (\bibinfo{year}{2014}), \bibinfo{pages}{529--607}.
\newblock


\bibitem[\protect\citeauthoryear{Stone and Harper}{Stone and Harper}{2006}]%
        {stoneharper}
\bibfield{author}{\bibinfo{person}{Christopher~A. Stone} {and}
  \bibinfo{person}{Robert Harper}.} \bibinfo{year}{2006}\natexlab{}.
\newblock \showarticletitle{Extensional equivalence and singleton types}.
\newblock \bibinfo{journal}{\emph{{ACM} Trans. Comput. Log.}}
  \bibinfo{volume}{7}, \bibinfo{number}{4} (\bibinfo{year}{2006}),
  \bibinfo{pages}{676--722}.
\newblock


\bibitem[\protect\citeauthoryear{Wang and Rompf}{Wang and Rompf}{2017}]%
        {normalization}
\bibfield{author}{\bibinfo{person}{Fei Wang} {and} \bibinfo{person}{Tiark
  Rompf}.} \bibinfo{year}{2017}\natexlab{}.
\newblock \showarticletitle{Towards Strong Normalization for Dependent Object
  Types {(DOT)}}. In \bibinfo{booktitle}{\emph{31st European Conference on
  Object-Oriented Programming, {ECOOP} 2017, June 19-23, 2017, Barcelona,
  Spain}}. \bibinfo{pages}{27:1--27:25}.
\newblock


\bibitem[\protect\citeauthoryear{Wright and Felleisen}{Wright and
  Felleisen}{1994}]%
        {felleisen}
\bibfield{author}{\bibinfo{person}{Andrew~K. Wright} {and}
  \bibinfo{person}{Matthias Felleisen}.} \bibinfo{year}{1994}\natexlab{}.
\newblock \showarticletitle{A Syntactic Approach to Type Soundness}.
\newblock \bibinfo{journal}{\emph{Inf. Comput.}} \bibinfo{volume}{115},
  \bibinfo{number}{1} (\bibinfo{year}{1994}), \bibinfo{pages}{38--94}.
\newblock


\end{thebibliography}

\appendix
\setlength{\multicolsep}{12pt}
\section{Appendix}

\subsection{Proof Recipe of \pdot}

In this section, we present an overview of the proof recipe (see Section~\ref{sec:proofrecipe}) and the auxiliary typing judgments that are involved in it.
The purpose of the proof recipe is to
reason about the most precise inert type assigned to a path by the typing environment,
given the path's general type.
The proof recipe works by performing a sequence of type transformations
starting with General typing and ending in Elimination typing.

The following diagram summarizes the typing relations that
participate in the proof recipe:
\begin{align*}
{\scriptsize
\text{General } (\vdash) \rightarrow
\text{Tight } (\turnstileTight)\rightarrow
\new{\text{{Introduction-$qp$ }} (\turnstileRepl)} \rightarrow
\text{Introduction-$pq$ }(\turnstileInvertible) \rightarrow
\new{\text{Elim-III }(\turnstilePrecThree)} \rightarrow
\new{\text{Elim-II }(\turnstilePrecTwo)} \rightarrow
\text{Elim-I }(\turnstilePrecOne)}
\end{align*}
In addition, Table~\ref{tab:typings} shows which typing rules of \pdot are handled by each of the
auxiliary typing relations.
The exact definitions of the typing relations are presented in
Figures~\ref{fig:tighttyping} to~\ref{fig:prectyping}.
\begin{table}[]
{\eqfontsize
\begin{tabular}{@{}llll@{}}
    \toprule
    \multicolumn{2}{l}{Relation} & Type rules & Inlined subtyping rules (\rn{Sub}+\dots) \\ \midrule
    \multirow{3}{*}{Elimination} & $\turnstilePrecOne$ & \rn{Var}, \new{\rn{Fld-E}}, \rn{Rec-E} & \rn{And$_1$}-$<:$, \rn{And$_2$}-$<:$ \\
    &$\turnstilePrecTwo$ & \new{\rn{Sngl-E}} &  \\
    &$\turnstilePrecThree$ & \new{\rn{Sngl-Trans}} &  \\\hline
    \multirow{2}{*}{Introduction}&$\turnstileInvertible$ & \&-I & \new{\rn{Sngl$_{pq}$-$<:$}}, $<:$-\rn{And}, $\rn{Top}$, \rn{All-$<:$-All}, \rn{Fld-$<:$-Fld}, \rn{Typ-$<:$-Typ} \\
    &$\turnstileRepl$ & \rn{\&-I}, \new{\rn{Fld-I}}, \rn{Rec}-I & \new{\rn{Sngl$_{qp}$-$<:$}}, $<:$-\rn{And}, $<:$-\rn{Sel} \\\hline
    Tight&$\turnstileTight$ & all & all, tight versions of \rn{Sel} rules \\
    General&$\vdash$ & all & all\\
     \bottomrule
\end{tabular}
}
\caption{Auxiliary typing relations that make up the proof recipe of \pdot}
\label{tab:typings}
\end{table}
\begin{wide-rules}

\textbf{Tight term typing}

\begin{multicols}{2}

\infrule[Var$_{\#}$]
  {\G(x)=T}
  {\typTightDft x T}
  
\infrule[All-I$_{\#}$]
  {\typ {\extendG x T} t U
    \andalso
    x\notin\fv T}
  {\typTightDft{\tLambda x T t}{\tForall x T U}}

\infrule[All-E$_{\#}$]
  {\typTightDft {\new p} {\tForall z S T}
    \andalso
    \typTightDft {\new q} S}
  {\typTightDft {\new{p\, q}} {\tSubst z {\new q} T}}

\infrule[\{\}-I$_{\#}$]
  {\dtypN x {\extendG x T} d T}
  {\typTightDft {\tNew x T d} {\tRec x T}}
  
\infrule[Fld-E$_{\#}$]
  {\typTightDft {\new p} {\tFldDec a T}}
  {\typTightDft {\new p.a} T}
  
\newruletrue
\infrule[Fld-I$_{\#}$]
{\typTightDft {p.a} T}
{\typTightDft p {\tFldDec a T}}
\newrulefalse

\infrule[Let$_{\#}$]
  {\typTightDft t T
      \\
    \typ {\extendG x T} u U
    \andalso
    x\notin\fv U}
  {\typTightDft {\ttLet x t u} U}

\newruletrue

\infrule[Sngl-Trans$_{\#}$]
{\typTightDft p {\single q} \andalso \typDft q T}
{\typTightDft p T}

\newruletrue
\infrule[Sngl-E$_{\#}$]
{\typTightDft p {\single q} \andalso
    \typeableTight {q.a}}
{\typTightDft {p.a} {\single {q.a}}}
\newrulefalse

\infrule[Rec-I$_{\#}$]
  {\typTightDft {\new p} {\new{\tSubst x p T}}}
  {\typTightDft {\new p} {\tRec x T}}

\infrule[Rec-E$_{\#}$]
  {\typTightDft {\new p} {\tRec x T}}
  {\typTightDft {\new p} {\new{\tSubst x p T}}}

\infrule[\&-I$_{\#}$]
  {\typTightDft {\new p} T
    \andalso
    \typTightDft {\new p} U}
  {\typTightDft {\new p} {\tAnd T U}}

\infrule[Sub$_{\#}$]
  {\typTightDft t T
    \andalso
    \subDft T U}
  {\typTightDft t U}

\end{multicols}

\textbf{Tight subtyping}
\begin{multicols}{2}
    
\infax[Top$_{\#}$]
  {\subTightDft T \top}

\infax[Bot$_{\#}$]
  {\subTightDft \bot T}

\infax[Refl$_{\#}$]
  {\subTightDft T T}

\infrule[Trans$_{\#}$]
  {\subTightDft S T
    \andalso
    \subTightDft T U}
  {\subTightDft S U}

\infax[And$_1$-$<:$$_{\#}$]
  {\subTightDft {\tAnd T U} T}

\infax[And$_2$-$<:$$_{\#}$]
  {\subTightDft {\tAnd T U} U}

\infrule[$<:$-And$_{\#}$]
  {\subTightDft S T
    \andalso
    \subTightDft S U}
  {\subTightDft S {\tAnd T U}}
  
\infrule[$<:$-Sel$_{\#}$]
  {\new{\typPrecThreeDft p {\tTypeDec A S S}}}
  {\subTightDft S {{\new p}.A}}

\infrule[Sel-$<:$$_{\#}$]
  {\new{\typPrecThreeDft p {\tTypeDec A S S}}}
  {\subTightDft {{\new p}.A} S}
  
\newruletrue
\infrule[Sngl$_{pq}$-$<:$$_{\#}$]
  {\typPrecThreeDft p {\single q} \andalso \typeablePrecTwo q}
  {\subTightDft T {\repl p q T}}

\infrule[Sngl$_{qp}$-$<:$$_{\#}$]
  {\typPrecThreeDft p {\single q} \andalso \typeablePrecTwo q}
  {\subTightDft T {\repl q p T}}
  
\newrulefalse

\infrule[Fld-$<:$-Fld$_{\#}$]
  {\subTightDft T U}
  {\subTightDft {\tFldDec a T} {\tFldDec a U}}

\infrule[Typ-$<:$-Typ$_{\#}$]
  {\subTightDft {S_2} {S_1}
    \andalso
    \subTightDft {T_1} {T_2}}
  {\subTightDft {\tTypeDec A {S_1} {T_1}} {\tTypeDec A {S_2} {T_2}}}

\infrule[All-$<:$-All$_{\#}$]
  {\subTightDft {S_2} {S_1}
    \\
    \sub {\extendG x {S_2}} {T_1} {T_2}}
  {\subTightDft {\tForall x {S_1} {T_1}} {\tForall x {S_2} {T_2}}}

  \end{multicols}
    
  \caption{Tight typing}
\label{fig:tighttyping}

\end{wide-rules}

The proof recipe starts by translating general into \textit{tight} typing,
denoted $\typTightDft{p}{T}$, which is
equivalent to general typing but does not admit the introduction
of new subtyping relationships. Tight typing is presented
in Figure~\ref{fig:tighttyping}.
The following lemma states that in an inert context,
the general type of a term is the same as its tight type:

\begin{lemma}[$\vdash$ to $\turnstileTight$]
    If $\typDft t T$ and $\G$ is inert then $\typTightDft{t}{T}$.
\end{lemma}

Each of the remaining proof-recipe typing relations comes in two versions:
for paths and for values. We present only the versions for paths;
the formulations of the lemmas for values are similar.
\begin{wide-rules}

\textbf{Introduction-$qp$ path typing}
\begin{multicols}{2}
\newruletrue

\infrule[Path$_{qp}$]
  {\tptDft p T}
  {\typrDft p T}

\infrule[And$_{qp}$]
  {\typrDft p S \andalso \typrDft p T}
  {\typrDft p {\tAnd S T}}

\infrule[Rec-I$_{qp}$]
{\typrDft {\new p} {\new{\tSubst x p T}}}
{\typrDft {\new p} {\tRec x T}}

\infrule[Fld-I$_{qp}$]
{\typrDft {p.a} T}
{\typrDft {p} {\tFldDec a T}}
  
\infrule[Typ-Sel$_{qp}$]
	{\typrDft p T \\
	 \typPrecDft q {\tTypeDec A T T}}
	{\typrDft p {q.A}}

\infrule[Sngl-Rec$_{qp}$]
{\typrDft r {\tRec x T} \\ 
	\typPrecDft p {\single q}  \andalso \typeablePrecTwo q}
{\typrDft r {\repl q p {\tRec x T}}}

\infrule[Sngl-Sel$_{qp}$]
{\typrDft r {r'.A} \\ 
	\typPrecDft p {\single q} \andalso \typeablePrecTwo q}
{\typrDft r {\repl q p {r'.A}}}

\infrule[Sngl-Sngl$_{qp}$]
{\typrDft r {\single {r'}} \\ 
	\typPrecDft p {\single q} \andalso \typeablePrecTwo q}
{\typrDft r {\repl q p {\single {r'}}}}

\newrulefalse

\end{multicols}

\textbf{Introduction$_{qp}$ value typing}
\begin{multicols}{2}
	\newruletrue
	
	\infrule[Path$_{qp}$-$v$]
	{\tptDft v T}
	{\typrDft v T}
	
	\infrule[And$_{qp}$-$v$]
	{\typrDft v S \andalso \typrDft p T}
	{\typrDft v {\tAnd S T}}
	
	\infrule[Typ-Sel$_{qp}$-$v$]
	{\typrDft v T \\
		\typPrecDft q {\tTypeDec A T T}}
	{\typrDft v {q.A}}
	
	\infrule[Sngl-Rec$_{qp}$-$v$]
	{\typrDft v {\tRec x T} \\
		\typPrecDft p {\single q} \andalso \typeablePrecTwo q}
	{\typrDft v {\repl q p {\tRec x T}}}
	
	\infrule[Rec-I$_{qp}$-$v$]
	{\typrDft v {\new{\tSubst x p T}}}
	{\typrDft v {\tRec x T}}
	
\infrule[Sngl-Sel$_{qp}$-$v$]
{\typrDft v {r'.A} \\ 
	\typPrecDft p {\single q} \andalso \typeablePrecTwo q}
{\typrDft v {\repl q p {r'.A}}}
\newrulefalse
		
\end{multicols}

  \caption{Introduction$_{qp}$ Typing}
  \label{fig:repltyping}

\end{wide-rules}
After obtaining a tight type of a path, we can convert it into
an Introduction-$qp$ type:

\begin{lemma}[$\turnstileTight$ to $\turnstileRepl$]
    If $\typTightDft p T$ and $\G$ is inert then
    $\typrDft p T$.
\end{lemma}
\noindent
Introduction-$qp$ typing,
denoted $\typrDft p T$,
contains introduction rules (such as recursion introduction)
and inlined versions of the \rn{Sngl$_{qp}$} subtyping rule
(see Figure~\ref{fig:repltyping}).
\begin{wide-rules}

\textbf{Introduction-$pq$ path typing}
\begin{multicols}{2}

\infrule[Path$_{pq}$]
  {\new{\typPrecThreeDft p T}}
  {\tptDft {\new p} T}
  
\infrule[Fld-$<:$$_{pq}$]
  {\tptDft {\new p} {\tFldDec a S} \andalso
   \subTightDft S T}
  {\tptDft {\new p} {\tFldDec a T}}
  
\infrule[Typ$_{pq}$]
  {\tptDft {\new p} {\tTypeDec A T U} \\
   \subTightDft {T'} T \andalso
   \subTightDft U {U'}}
  {\tptDft {\new p} {\tTypeDec A {T'} {U'}}}

\infrule[All$_{pq}$]
  {\tptDft {\new p} {\tForall x S T} \\
   \subTightDft {S'} S \andalso
   \sub {\extendG x {S'}} T {T'}}
  {\tptDft {\new p} {\tForall x {S'} {T'}}}

\infrule[And$_{pq}$]
  {\tptDft {\new p} S \andalso \tptDft {\new p} T}
  {\tptDft {\new p} {\tAnd S T}}

\infrule[Top$_{pq}$]
  {\tptDft {\new p} T}
  {\tptDft {\new p} \top}

\newruletrue

\infrule[Sngl-Rec$_{pq}$]
	{\tptDft r {\tRec x T} \\ 
	 \typPrecDft p {\single q}  \andalso \typeablePrecTwo q}
	{\tptDft r {\repl p q {\tRec x T}}}
	
\infrule[Sngl-Sel$_{pq}$]
	{\tptDft r {r'.A} \\ 
	 \typPrecDft p {\single q} \andalso \typeablePrecTwo q}
	{\tptDft r {\repl p q {r'.A}}}

\infrule[Sngl-Sngl$_{pq}$]
	{\tptDft r {\single {r'}} \\ 
	 \typPrecDft p {\single q} \andalso \typeablePrecTwo q}
	{\tptDft r {\repl p q {\single {r'}}}}

\newrulefalse

\end{multicols}

\textbf{Introduction-$pq$ value typing}
\begin{multicols}{2}
\infrule[Path-$v$$_{pq}$]
  {\typPrecDft v T}
  {\tptDft v T}
  
\infrule[All-$v$$_{pq}$]
  {\tptDft v {\tForall x S T} \\
   \subTightDft {S'} S \andalso
   \sub {\extendG x {S'}} T {T'}}
  {\typDft v {\tForall x {S'} {T'}}}

\infrule[And-$v$$_{pq}$]
  {\tptDft v S \andalso \tptDft v T}
  {\tptDft v {\tAnd S T}}
  
\infrule[Top-$v$$_{pq}$]
  {\tptDft v T}
  {\tptDft v \top}
  
\newruletrue
\infrule[Rec-Sngl-$v$$_{pq}$]
{\tptDft v {\tRec x T} \\ 
	\typPrecDft p {\single q}  \andalso \typeablePrecTwo q}
{\tptDft v {\repl p q {\tRec x T}}}  
\newrulefalse
\end{multicols}
    
  \caption{Introduction-$pq$ Typing}
  \label{fig:invtyping}

\end{wide-rules}

The remaining proof recipe lemmas are specialized to function and
singleton types, and we will only present the ones for function types.
The following lemma establishes that if a path $p$ has a function type
in Introduction-$qp$ typing then it has the same type in Introduction-$pq$ typing:%
\footnote{The versions of Lemma~\ref{lem:qptopq}
    for singleton and object types are more complicated:
    the relationship between a path's Introduction-$qp$ as Introduction-$pq$ type
    will be equivalence (i.e. possible replacements of aliased paths)
    rather than equality.}

\begin{lemma}[$\turnstileRepl$ to $\turnstileInvertible$ ($\forall$)]
    \label{lem:qptopq}
    If $\typrDft p {\tForall x T U}$ and $\G$ is inert then
    $\tptDft p {\tForall x T U}$.
\end{lemma}
\noindent
\begin{wide-rules}

\textbf{Elimination typing for values}

\begin{multicols}{2}

\infrule[All-I$_\precOne$]
  {\typ {\extendG x T} t U
    \andalso
    x\notin\fv T}
  {\typPrecDft{\tLambda x T t}{\tForall x T U}}

\infrule[\{\}-I$_\precOne$]
  {\dtypN x {\extendG x T} d T}
  {\typPrecDft {\tNew x T d} {\tRec x T}}
\end{multicols}

\textbf{Elimination typing for paths}\\
$\prectypingonesymbol$

\begin{multicols}{2}
\infrule[Var$_\precOne$]
  {\G(x)=T}
  {\typPrecDft x T}

\infrule[Rec-E$_\precOne$]
  {\typPrecDft {\new p} {\tRec z T}}
  {\typPrecDft {\new p} {\tSubst z {\new p} T}}
  
\infrule[And$_1$-E$_\precOne$]
  {\typPrecDft {\new p} {\tAnd T U}}
  {\typPrecDft {\new p} T}

\infrule[And$_2$-E$_\precOne$]
  {\typPrecDft {\new p} {\tAnd T U}}
  {\typPrecDft {\new p} U}
  
\newruletrue

\infrule[Fld-E$_\precOne$]
  {\typPrecDft p {\tFldDec a T}}
  {\typPrecDft {p.a} T}
  
\newrulefalse

\end{multicols}

$\new{\prectypingtwosymbol}$
\begin{multicols}{2}

\newruletrue
\infrule[Path$_\precTwo$]
  {\typPrecDft p T}
  {\typPrecTwoDft p T}

\newruletrue  
\infrule[Sngl-E$_\precTwo$]
  {\typPrecTwoDft p {\single q} \andalso
   \typeablePrecTwo {q.a}}
  {\typPrecTwoDft {p.a} {\single{q.a}}}
  
\end{multicols}

$\new{\prectypingthreesymbol}$
\begin{multicols}{2}

\newruletrue
\infrule[Path$_\precThree$]
  {\typPrecTwoDft p T}
  {\typPrecThreeDft p T}

\newruletrue  
\infrule[Sngl-Trans$_\precThree$]
	{\typPrecTwoDft p {\single q}
		\andalso
		\typPrecThreeDft q U}
	{\typPrecThreeDft p U}
  
\newrulefalse 

\end{multicols}

  \caption{Elimination Typing Rules}
  \label{fig:prectyping}

\end{wide-rules}

Introduction-$pq$ typing,
denoted $\tptDft p T$,
contains introduction rules
and inlined versions of the \rn{Sngl$_{pq}$} subtyping rule
(see Figure~\ref{fig:invtyping}).

To convert a function type from Introduction-$pq$ typing into
Elimination-III typing, we use the following lemma.

\begin{lemma}[$\turnstileInvertible$ to $\turnstilePrecThree$ ($\forall$)]
    If $\tptDft p {\tForall x T U}$ and $\G$ is inert then
    there exist types $T'$ and $U'$ such that
    $\subTightDft{T}{T'}$,
    $\subTight {\extendG x T} {U'} U$, and
    $\typPrecThreeDft p {\tForall x {T'} {U'}}$.
\end{lemma}

If $p$ has type $T$ in Elimination-III typing,
denoted $\typPrecThreeDft p T$,
we know that either the environment directly assigns $p$ the type $T$
or that $p$ is assigned a singleton type $q$, and by recursively following
path aliases starting with $q$ we eventually arrive at $T$. More precisely,
if $\typPrecThreeDft p T$, then one of the following is true:
\begin{enumerate}[1)]
    \item $\typPrecDft p T$, i.e. $T$ is the most precise type that
        $\G$ assigns to $p$ (modulo possible recursion and intersection
        elimination), or
    \item $p=p'.\overline b$ (i.e. $p'$ is a prefix of $p$),
        $\typPrecDft {p'} {\single q}$
        (i.e. $\typPrecTwoDft {p'.\overline b} 
            {\single{q.\overline b}}$),
         and either
         \begin{itemize}
            \item $T=\single{q.\overline b}$, or 
            \item $\typPrecThreeDft {q.\overline b} T$.
        \end{itemize}
\end{enumerate}
The Elimination typing rules are presented in Figure~\ref{fig:prectyping}.
For the proof recipe, we do not need to further convert $p$'s type
into Elimination-II and -I typings. However, these typing relations are
needed to prove the lemmas of the proof recipe.

\section{Correspondence with Artifact}
\label{pdot-type-safety-proof}

This section contains a correspondence guide between the \pdot type-safety
proof as presented in Section~\ref{sec:proof} and its Coq formalization
which is presented in the accompanying artifact
 and can be also found under
\begin{center}
    \href{https://git.io/dotpaths}{https://git.io/dotpaths}
\end{center}

\newcommand{\urlprefix}{https://amaurremi.github.io/dot-calculus/src/extensions/paths/doc/}

\subsection{Compiling the Proof}\label{compiling-the-proof}

\emph{System requirements}:

\begin{itemize}
    \tightlist
    \item
    make
    \item
    an installation of \href{https://coq.inria.fr/opam-using.html}{Coq
        8.9.0}, preferably through \href{https://opam.ocaml.org/}{opam}
    \item
    the \href{https://gitlab.inria.fr/charguer/tlc}{TLC} library which can
    be installed through
\end{itemize}

\begin{verbatim}
opam repo add coq-released http://coq.inria.fr/opam/released
opam install -j4 coq-tlc
\end{verbatim}

To compile the proof navigate to the \code{paths} directory and run
\code{make}.

\subsection{Paper Correspondence}\label{sec:correspondence}

The pDOT calculus is formalized using the
locally nameless representation with cofinite quantification~\cite{ln}
in which free variables are
represented as named variables, and bound variables are represented as
de Bruijn indices.

We include the
\href{\urlprefix Sequences.html}{Sequences}
library by Xavier Leroy into our
development to reason about the reflexive, transitive closure of binary
relations.

\newcommand{\defwidth}{0.11\columnwidth}
\newcommand{\inpapwidth}{0.09\columnwidth}
\newcommand{\filewidth}{0.11\columnwidth}
\newcommand{\papnotwidth}{0.15\columnwidth}
\newcommand{\proofnotwidth}{0.25\columnwidth}
\newcommand{\inproofwidth}{0.13\columnwidth}

\makeatletter
\def\fps@figure{htbp}
\makeatother

The correspondence between the paper and Coq formalization
is documented in Tables~\ref{tab:correspondence-of-definitions},
\ref{tab:correspondence-of-lemmas-and-theorems}, and~\ref{tab:correspondence-of-examples}.

\scriptsize
\begin{longtable}[h!]{@{}llllll@{}}
    \caption{Correspondence of Definitions}
    \label{tab:correspondence-of-definitions}\\
    \toprule
    \begin{minipage}[b]{\defwidth}\raggedright
        Definition\strut
    \end{minipage} & \begin{minipage}[b]{\inpapwidth}\raggedright
        In paper\strut
    \end{minipage} & \begin{minipage}[b]{\filewidth}\raggedright
        File\strut
    \end{minipage} & \begin{minipage}[b]{\papnotwidth}\raggedright
        Paper notations\strut
    \end{minipage} & \begin{minipage}[b]{\proofnotwidth}\raggedright
        Proof notations\strut
    \end{minipage} & \begin{minipage}[b]{\inproofwidth}\raggedright
        In proof\strut
    \end{minipage}\tabularnewline
    \midrule
    \endhead
    \begin{minipage}[t]{\defwidth}\raggedright
        Abstract Syntax\strut
    \end{minipage} & \begin{minipage}[t]{\inpapwidth}\raggedright
        Fig.~\ref{fig:synt} \strut
    \end{minipage} & \begin{minipage}[t]{\filewidth}\raggedright
        \href{\urlprefix Definitions.html}{Definitions.v}\strut
    \end{minipage} & \begin{minipage}[t]{\papnotwidth}\raggedright
        \strut
    \end{minipage} & \begin{minipage}[t]{\proofnotwidth}\raggedright
        \strut
    \end{minipage} & \begin{minipage}[t]{\filewidth}\raggedright
        \strut
    \end{minipage}\tabularnewline
    \begin{minipage}[t]{\defwidth}\raggedright
        - variable\strut
    \end{minipage} & \begin{minipage}[t]{\inpapwidth}\raggedright
        Fig.
        \ref{fig:synt}\strut
    \end{minipage} & \begin{minipage}[t]{\filewidth}\raggedright
        \href{\urlprefix Definitions.html}{Definitions.v}\strut
    \end{minipage} & \begin{minipage}[t]{\papnotwidth}\raggedright
        \strut
    \end{minipage} & \begin{minipage}[t]{\proofnotwidth}\raggedright
        \strut
    \end{minipage} & \begin{minipage}[t]{\filewidth}\raggedright
        \href{\urlprefix Definitions.html\#avar}{\texttt{avar}}\strut
    \end{minipage}\tabularnewline
    \begin{minipage}[t]{\defwidth}\raggedright
        - term member\strut
    \end{minipage} & \begin{minipage}[t]{\inpapwidth}\raggedright
        Fig.
        \ref{fig:synt}\strut
    \end{minipage} & \begin{minipage}[t]{\filewidth}\raggedright
        \href{\urlprefix Definitions.html}{Definitions.v}\strut
    \end{minipage} & \begin{minipage}[t]{\papnotwidth}\raggedright
        \strut
    \end{minipage} & \begin{minipage}[t]{\proofnotwidth}\raggedright
        \strut
    \end{minipage} & \begin{minipage}[t]{\filewidth}\raggedright
        \href{\urlprefix Definitions.html\#trm_label}{\texttt{trm\_label}}\strut
    \end{minipage}\tabularnewline
    \begin{minipage}[t]{\defwidth}\raggedright
        - type member\strut
    \end{minipage} & \begin{minipage}[t]{\inpapwidth}\raggedright
        Fig.
        \ref{fig:synt}\strut
    \end{minipage} & \begin{minipage}[t]{\filewidth}\raggedright
        \href{\urlprefix Definitions.html}{Definitions.v}\strut
    \end{minipage} & \begin{minipage}[t]{\papnotwidth}\raggedright
        \strut
    \end{minipage} & \begin{minipage}[t]{\proofnotwidth}\raggedright
        \strut
    \end{minipage} & \begin{minipage}[t]{\filewidth}\raggedright
        \href{\urlprefix Definitions.html\#typ_label}{\texttt{typ\_label}}\strut
    \end{minipage}\tabularnewline
    \begin{minipage}[t]{\defwidth}\raggedright
        - path\strut
    \end{minipage} & \begin{minipage}[t]{\inpapwidth}\raggedright
        Fig.
        \ref{fig:synt}\strut
    \end{minipage} & \begin{minipage}[t]{\filewidth}\raggedright
        \href{\urlprefix Definitions.html}{Definitions.v}\strut
    \end{minipage} & \begin{minipage}[t]{\papnotwidth}\raggedright
        $x.a.b.c$\\$p.a$\\$p.b$\strut
    \end{minipage} & \begin{minipage}[t]{\proofnotwidth}\raggedright
        \texttt{p\_sel\ x\ (c::b::a::nil)}\\\texttt{p$\cdot$a}\\\texttt{p$\cdot\cdot$b}\strut
    \end{minipage} & \begin{minipage}[t]{\filewidth}\raggedright
        \href{\urlprefix Definitions.html\#path}{\texttt{path}}\strut
    \end{minipage}\tabularnewline
    \begin{minipage}[t]{\defwidth}\raggedright
        - term\strut
    \end{minipage} & \begin{minipage}[t]{\inpapwidth}\raggedright
        Fig.
        \ref{fig:synt}\strut
    \end{minipage} & \begin{minipage}[t]{\filewidth}\raggedright
        \href{\urlprefix Definitions.html}{Definitions.v}\strut
    \end{minipage} & \begin{minipage}[t]{\papnotwidth}\raggedright
        \strut
    \end{minipage} & \begin{minipage}[t]{\proofnotwidth}\raggedright
        \strut
    \end{minipage} & \begin{minipage}[t]{\filewidth}\raggedright
        \href{\urlprefix Definitions.html\#trm}{\texttt{trm}}\strut
    \end{minipage}\tabularnewline
    \begin{minipage}[t]{\defwidth}\raggedright
        - stable term\strut
    \end{minipage} & \begin{minipage}[t]{\inpapwidth}\raggedright
        Fig.
        \ref{fig:synt}\strut
    \end{minipage} & \begin{minipage}[t]{\filewidth}\raggedright
        \href{\urlprefix Definitions.html}{Definitions.v}\strut
    \end{minipage} & \begin{minipage}[t]{\papnotwidth}\raggedright
        \strut
    \end{minipage} & \begin{minipage}[t]{\proofnotwidth}\raggedright
        \strut
    \end{minipage} & \begin{minipage}[t]{\filewidth}\raggedright
        \href{\urlprefix Definitions.html\#def_rhs}{\texttt{def\_rhs}}\strut
    \end{minipage}\tabularnewline
    \begin{minipage}[t]{\defwidth}\raggedright
        - value\strut
    \end{minipage} & \begin{minipage}[t]{\inpapwidth}\raggedright
        Fig.
        \ref{fig:synt}\strut
    \end{minipage} & \begin{minipage}[t]{\filewidth}\raggedright
        \href{\urlprefix Definitions.html}{Definitions.v}\strut
    \end{minipage} & \begin{minipage}[t]{\papnotwidth}\raggedright
        $\tNew x T d$\\$\tLambda x T t$\strut
    \end{minipage} & \begin{minipage}[t]{\proofnotwidth}\raggedright
        \texttt{$\nu$(T)d}\\ \texttt{$\lambda$(T)t}\strut
    \end{minipage} & \begin{minipage}[t]{\filewidth}\raggedright
        \href{\urlprefix Definitions.html\#val}{\texttt{val}}\strut
    \end{minipage}\tabularnewline
    \begin{minipage}[t]{\defwidth}\raggedright
        - definition\strut
    \end{minipage} & \begin{minipage}[t]{\inpapwidth}\raggedright
        Fig.
        \ref{fig:synt}\strut
    \end{minipage} & \begin{minipage}[t]{\filewidth}\raggedright
        \href{\urlprefix Definitions.html}{Definitions.v}\strut
    \end{minipage} & \begin{minipage}[t]{\papnotwidth}\raggedright
        $\set{a=t}$\\$\set{A=T}$\strut
    \end{minipage} & \begin{minipage}[t]{\proofnotwidth}\raggedright
        \texttt{\{a\ :=\ t\}}\\ \texttt{\{A\ :=\ T\}}\strut
    \end{minipage} & \begin{minipage}[t]{\filewidth}\raggedright
        \href{\urlprefix Definitions.html\#def}{\texttt{def}}\strut
    \end{minipage}\tabularnewline
    \begin{minipage}[t]{\defwidth}\raggedright
        - type\strut
    \end{minipage} & \begin{minipage}[t]{\inpapwidth}\raggedright
        Fig.
        \ref{fig:synt}\strut
    \end{minipage} & \begin{minipage}[t]{\filewidth}\raggedright
        \href{\urlprefix Definitions.html}{Definitions.v}\strut
    \end{minipage} & \begin{minipage}[t]{\papnotwidth}\raggedright
        $\tFldDec a T$\\
        $\tTypeDec A T U$\\
        $\tForall x T U$\\
        $p.A$\\
        $\single p$\\
        $\tRec x T$\\
        $\tAnd T U$\\
        $\top$\\
        $\bot$\strut
    \end{minipage} & \begin{minipage}[t]{\proofnotwidth}\raggedright
        \texttt{\{a\ :\ T\}}\\
        \texttt{\{A\ $>:$\ T\ $<:$\ U\}}\\
        \texttt{$\forall$(T)U} \\
        \texttt{p$\downarrow$A} \\
        \texttt{\{\{p\}\}}\\
        \texttt{$\mu$(T)}\\
        \texttt{T\ $\wedge$\ U} \\
        \texttt{$\top$} \\
        \texttt{$\bot$}\strut
    \end{minipage} & \begin{minipage}[t]{\filewidth}\raggedright
        \href{\urlprefix Definitions.html\#typ}{\texttt{typ}}\strut
    \end{minipage}\tabularnewline
    \begin{minipage}[t]{\defwidth}\raggedright
        Type System\strut
    \end{minipage} & \begin{minipage}[t]{\inpapwidth}\raggedright
    \end{minipage} & \begin{minipage}[t]{\filewidth}\raggedright

    \end{minipage} & \begin{minipage}[t]{\papnotwidth}\raggedright
        \strut
    \end{minipage} & \begin{minipage}[t]{\proofnotwidth}\raggedright
        \strut
    \end{minipage} & \begin{minipage}[t]{\filewidth}\raggedright
        \strut
    \end{minipage}\tabularnewline
    \begin{minipage}[t]{\defwidth}\raggedright
        - term typing\strut
    \end{minipage} & \begin{minipage}[t]{\inpapwidth}\raggedright
        Fig. \ref{fig:typing}\strut
    \end{minipage} & \begin{minipage}[t]{\filewidth}\raggedright
        \href{\urlprefix Definitions.html}{Definitions.v}\strut
    \end{minipage} & \begin{minipage}[t]{\papnotwidth}\raggedright
        $\typDft t T$\strut
    \end{minipage} & \begin{minipage}[t]{\proofnotwidth}\raggedright
        \texttt{$\G$\ $\vdash$\ t\ :\ T}\strut
    \end{minipage} & \begin{minipage}[t]{\filewidth}\raggedright
        \href{\urlprefix Definitions.html\#ty_trm}{\texttt{ty\_trm}}\strut
    \end{minipage}\tabularnewline
    \begin{minipage}[t]{\defwidth}\raggedright
        - definition typing\strut
    \end{minipage} & \begin{minipage}[t]{\inpapwidth}\raggedright
        Fig.
        \ref{fig:typing}\strut
    \end{minipage} & \begin{minipage}[t]{\filewidth}\raggedright
        \href{\urlprefix Definitions.html}{Definitions.v}\strut
    \end{minipage} & \begin{minipage}[t]{\papnotwidth}\raggedright
        $\dtypDft d T$\strut
    \end{minipage} & \begin{minipage}[t]{\proofnotwidth}\raggedright
        \texttt{x;\ bs;\ $\G$\ $\vdash$\ d\ :\ T} \\(single definition)\\
        \texttt{x;\ bs;\ $\G$\ $\vdash$\ d\ ::\ T}\\ (multiple definitions)\\ Here,
        $p$=\texttt{x.bs}, i.e.~\texttt{x} is $p$'s receiver, and \texttt{bs} are
        $p$'s fields in reverse order\strut
    \end{minipage} & \begin{minipage}[t]{\filewidth}\raggedright
        \href{\urlprefix Definitions.html\#ty_def}{\texttt{ty\_def}}
        \href{\urlprefix Definitions.html\#ty_defs}{\texttt{ty\_defs}}\strut
    \end{minipage}\tabularnewline
    \begin{minipage}[t]{\defwidth}\raggedright
        - tight bounds\strut
    \end{minipage} & \begin{minipage}[t]{\inpapwidth}\raggedright
        Fig.
        \ref{fig:typing}\strut
    \end{minipage} & \begin{minipage}[t]{\filewidth}\raggedright
        \href{\urlprefix Definitions.html}{Definitions.v}\strut
    \end{minipage} & \begin{minipage}[t]{\papnotwidth}\raggedright
        \strut
    \end{minipage} & \begin{minipage}[t]{\proofnotwidth}\raggedright
        \strut
    \end{minipage} & \begin{minipage}[t]{\filewidth}\raggedright
        \href{\urlprefix Definitions.html\#tight_bounds}{\texttt{tight\_bounds}}\strut
    \end{minipage}\tabularnewline
    \begin{minipage}[t]{\defwidth}\raggedright
        - subtyping\strut
    \end{minipage} & \begin{minipage}[t]{\inpapwidth}\raggedright
        Fig. \ref{fig:typing}
    \end{minipage} & \begin{minipage}[t]{\filewidth}\raggedright
        \href{\urlprefix Definitions.html}{Definitions.v}\strut
    \end{minipage} & \begin{minipage}[t]{\papnotwidth}\raggedright
        $\subDft T U$\strut
    \end{minipage} & \begin{minipage}[t]{\proofnotwidth}\raggedright
        \texttt{$\G$\ $\vdash$\ T\ \textless{}:\ U}\strut
    \end{minipage} & \begin{minipage}[t]{\filewidth}\raggedright
        \href{\urlprefix Definitions.html\#subtyp}{\texttt{subtyp}}\strut
    \end{minipage}\tabularnewline
    \begin{minipage}[t]{\defwidth}\raggedright
        Operational semantics\strut
    \end{minipage} & \begin{minipage}[t]{\inpapwidth}\raggedright
        Fig.
        \ref{fig:red2}\strut
    \end{minipage} & \begin{minipage}[t]{\filewidth}\raggedright
        \href{\urlprefix Reduction.html}{Reduction.v}\strut
    \end{minipage} & \begin{minipage}[t]{\papnotwidth}\raggedright
        $\reduction \sta t {\sta'} {t'}$\\
        ${\stDft t}\redt {\st {\sta'} {t'}}$\strut
    \end{minipage} & \begin{minipage}[t]{\proofnotwidth}\raggedright
        \texttt{($\sta$,\ t)\ $\longmapsto$\ ($\sta$\textquotesingle{},\ t\textquotesingle{})}\\
        \texttt{($\sta$,\ t)\ $\longmapsto$*\ ($\sta$\textquotesingle{},\ t\textquotesingle{})}\strut
    \end{minipage} & \begin{minipage}[t]{\filewidth}\raggedright
        \href{\urlprefix Reduction.html\#red}{\texttt{red}}\strut
    \end{minipage}\tabularnewline
    \begin{minipage}[t]{\defwidth}\raggedright
        Path lookup\strut
    \end{minipage} & \begin{minipage}[t]{\inpapwidth}\raggedright
        Fig.
        \ref{fig:lookup}\strut
    \end{minipage} & \begin{minipage}[t]{\filewidth}\raggedright
        \href{\urlprefix Lookup.html}{Lookup.v}\strut
    \end{minipage} & \begin{minipage}[t]{\papnotwidth}\raggedright
        $\lookupStepDft p s$
        \\
        $\lookupDft s {s'}$\strut
    \end{minipage} & \begin{minipage}[t]{\proofnotwidth}\raggedright
        \texttt{$\sta$\ $\llbracket$\ p\ $\leadsto$\ s\ $\rrbracket$}\\
        \texttt{$\sta$\ $\llbracket$\ s\ $\leadsto$*\ s\textquotesingle{}\ $\rrbracket$}\strut
    \end{minipage} & \begin{minipage}[t]{\filewidth}\raggedright
        \href{\urlprefix Lookup.html\#lookup_step}{\texttt{lookup\_step}}\strut
    \end{minipage}\tabularnewline
    \begin{minipage}[t]{\defwidth}\raggedright
        Extended reduction\strut
    \end{minipage} & \begin{minipage}[t]{\inpapwidth}\raggedright
        Sec. \ref{sec:proof}\strut
    \end{minipage} & \begin{minipage}[t]{\filewidth}\raggedright
        \href{\urlprefix Safety.html}{Safety.v}\strut
    \end{minipage} & \begin{minipage}[t]{\papnotwidth}\raggedright
        $\stDft t \redext {\st {\sta'} {t'}}$ \\
        $\stDft t \redextt {\st {\sta'} {t'}}$ \strut
    \end{minipage} & \begin{minipage}[t]{\proofnotwidth}\raggedright
        \texttt{($\sta$,\ t)$\redext$($\sta$\textquotesingle{},\ t\textquotesingle{})}\\
        \texttt{($\sta$,\ t)$\redextt$($\sta$\textquotesingle{},\ t\textquotesingle{})}\strut
    \end{minipage} & \begin{minipage}[t]{\filewidth}\raggedright
        \href{\urlprefix Safety.html\#extended_red}{\texttt{extended\_red}}\strut
    \end{minipage}\tabularnewline
    \begin{minipage}[t]{\defwidth}\raggedright
        Inert and record types\strut
    \end{minipage} & \begin{minipage}[t]{\inpapwidth}\raggedright
        Fig.
        \ref{fig:inert}\strut
    \end{minipage} & \begin{minipage}[t]{\filewidth}\raggedright
        \href{\urlprefix Definitions.html}{Definitions.v}\strut
    \end{minipage} & \begin{minipage}[t]{\papnotwidth}\raggedright
        inert T \\inert $\G$\strut
    \end{minipage} & \begin{minipage}[t]{\proofnotwidth}\raggedright
        \strut
    \end{minipage} & \begin{minipage}[t]{\filewidth}\raggedright
        \href{\urlprefix Definitions.html\#inert_typ}{\texttt{inert\_typ}}
        \href{\urlprefix Definitions.html\#inert}{\texttt{inert}}\strut
    \end{minipage}\tabularnewline
    \begin{minipage}[t]{\defwidth}\raggedright
        Well-formed environments\strut
    \end{minipage} & \begin{minipage}[t]{\inpapwidth}\raggedright
        Sec.
        \ref{sec:inertness}\strut
    \end{minipage} & \begin{minipage}[t]{\filewidth}\raggedright
        \href{\urlprefix PreciseTyping.html}{PreciseTyping.v}\strut
    \end{minipage} & \begin{minipage}[t]{\papnotwidth}\raggedright
        \strut
    \end{minipage} & \begin{minipage}[t]{\proofnotwidth}\raggedright
        \strut
    \end{minipage} & \begin{minipage}[t]{\filewidth}\raggedright
        \href{\urlprefix PreciseTyping.html\#wf}{\texttt{wf}}\strut
    \end{minipage}\tabularnewline
    \begin{minipage}[t]{\defwidth}\raggedright
        Correspondence between a value and type environment\strut
    \end{minipage} & \begin{minipage}[t]{\inpapwidth}\raggedright
        Sec. \ref{sec:proof}\strut
    \end{minipage} & \begin{minipage}[t]{\filewidth}\raggedright
        \href{\urlprefix Definitions.html}{Definitions.v}\strut
    \end{minipage} & \begin{minipage}[t]{\papnotwidth}\raggedright
        $\sta$: $\G$\strut
    \end{minipage} & \begin{minipage}[t]{\proofnotwidth}\raggedright
        \texttt{$\sta$\ $\fcolon$\ $\G$}\strut
    \end{minipage} & \begin{minipage}[t]{\filewidth}\raggedright
        \href{\urlprefix Definitions.html\#well_typed}{\texttt{well\_typed}}\strut
    \end{minipage}\tabularnewline
    \bottomrule
\end{longtable}

\normalsize

\begin{longtable}[]{@{}lll@{}}
    \caption{Correspondence of Lemmas and Theorems}
    \label{tab:correspondence-of-lemmas-and-theorems}\\
    \toprule
    Theorem & File & In proof\tabularnewline
    \midrule
    \endhead
    Theorem \ref{theorem}
    (Soundness) &
    \href{\urlprefix Safety.html}{Safety.v}
    &
    \href{\urlprefix Safety.html\#safety}{\texttt{safety}}\tabularnewline
    Theorem \ref{theorem-extended}
    (Extended Soundness) &
    \href{\urlprefix Safety.html}{Safety.v}
    &
    \href{\urlprefix Safety.html\#extended_safety}{\texttt{extended\_safety}}\tabularnewline
    Lemma \ref{lemma:progress}
    (Progress) &
    \href{\urlprefix Safety.html}{Safety.v}
    &
    \href{\urlprefix Safety.html\#progress}{\texttt{progress}}\tabularnewline
    Lemma \ref{lemma:preservation}
    (Preservation) &
    \href{\urlprefix Safety.html}{Safety.v}
    &
    \href{\urlprefix Safety.html\#preservation}{\texttt{preservation}}\tabularnewline
    Lemma \ref{lem:funcanforms} &
    \href{\urlprefix CanonicalForms.html}{CanonicalForms.v}
    &
    \href{\urlprefix CanonicalForms.html\#canonical_forms_fun}{\texttt{canonical\_forms\_fun}}\tabularnewline
    \bottomrule
\end{longtable}

\begin{longtable}[]{@{}lll@{}}
    \caption{Correspondence of Examples}
    \label{tab:correspondence-of-examples}\\
    \toprule
    Example & In paper & File\tabularnewline
    \midrule
    \endhead
    List example & Figure
    \ref{fig:examples} a &
    \href{\urlprefix ListExample.html}{ListExample.v}\tabularnewline
    Compiler example & Figure
    \ref{fig:examples} b&
    \href{\urlprefix CompilerExample.html}{CompilerExample.v}\tabularnewline
    Singleton type example & Figure
    \ref{fig:examples} c &
    \href{\urlprefix SingletonTypeExample.html}{SingletonTypeExample.v}\tabularnewline
    \bottomrule
\end{longtable}

\newpage \subsection{Proof Organization}\label{proof-organization}

\hypertarget{safety-proof}{%
    \subsubsection{Safety Proof}\label{safety-proof}}

The Coq proof is split up into the following modules:
\begin{itemize}
    \item 
\emph{\href{\urlprefix Definitions.html}{Definitions.v}}:
Definitions of pDOT's abstract syntax and type system.\item 
\emph{\href{\urlprefix Reduction.html}{Reduction.v}}:
Normal forms and the operational semantics of pDOT.\item 
\emph{\href{\urlprefix Safety.html}{Safety.v}}:
\emph{Final safety theorem} through Progress and Preservation.
\item 
\href{\urlprefix Lookup.html}{Lookup.v}:
Definition of path lookup and properties of lookup.\item 
\href{\urlprefix Binding.html}{Binding.v}:
Lemmas related to opening and variable binding.\item 
\href{\urlprefix SubEnvironments.html}{SubEnvironments.v}:
Lemmas related to subenvironments.\item 
\href{\urlprefix Weakening.html}{Weakening.v}:
Weakening Lemma.\item 
\href{\urlprefix RecordAndInertTypes.html}{RecordAndInertTypes.v}:
Lemmas related to record and inert types.\item 
\href{\urlprefix Replacement.html}{Replacement.v}:
Properties of equivalent types.\item 
\href{\urlprefix Narrowing.html}{Narrowing.v}:
Narrowing Lemma.\item 
\href{\urlprefix PreciseFlow.html}{PreciseFlow.v}
and
\href{\urlprefix PreciseTyping.html}{PreciseTyping.v}:
Lemmas related to elimination typing. In particular, reasons about the
possible precise types that a path can have in an inert environment.\item 
\href{\urlprefix TightTyping.html}{TightTyping.v}:
Defines tight typing and subtyping.\item 
\href{\urlprefix Substitution.html}{Substitution.v}:
Proves the Substitution Lemma.\item 
\href{\urlprefix InvertibleTyping.html}{InvertibleTyping.v}
and
\href{\urlprefix ReplacementTyping.html}{ReplacementTyping.v}:
Lemmas related to introduction typing.\item 
\href{\urlprefix GeneralToTight.html}{GeneralToTight.v}:
Proves that in an inert context, general typing implies tight typing.\item 
\href{\urlprefix CanonicalForms.html}{CanonicalForms.v}:
Canonical Forms Lemma.\item 
\href{\urlprefix Sequences.html}{Sequences.v}:
A library of relation operators by Xavier Leroy.
\end{itemize}

\hypertarget{examples}{%
    \subsubsection{Examples}\label{examples}}

\begin{itemize}
    \tightlist
    \item
    \href{\urlprefix CompilerExample.html}{CompilerExample.v}:
    The dotty-compiler example that contains paths of length greater than
    one.
    \item
    \href{\urlprefix ListExample.html}{ListExample.v}:
    A covariant-list implementation.
    \item
    \href{\urlprefix SingletonTypeExample.html}{SingletonTypeExample.v}:
    Method chaining through singleton types.
    \item
    \href{\urlprefix ExampleTactics.html}{ExampleTactics.v}:
    Helper tactics to prove the above examples.
\end{itemize}

\begin{figure}
    \centering
    \includegraphics{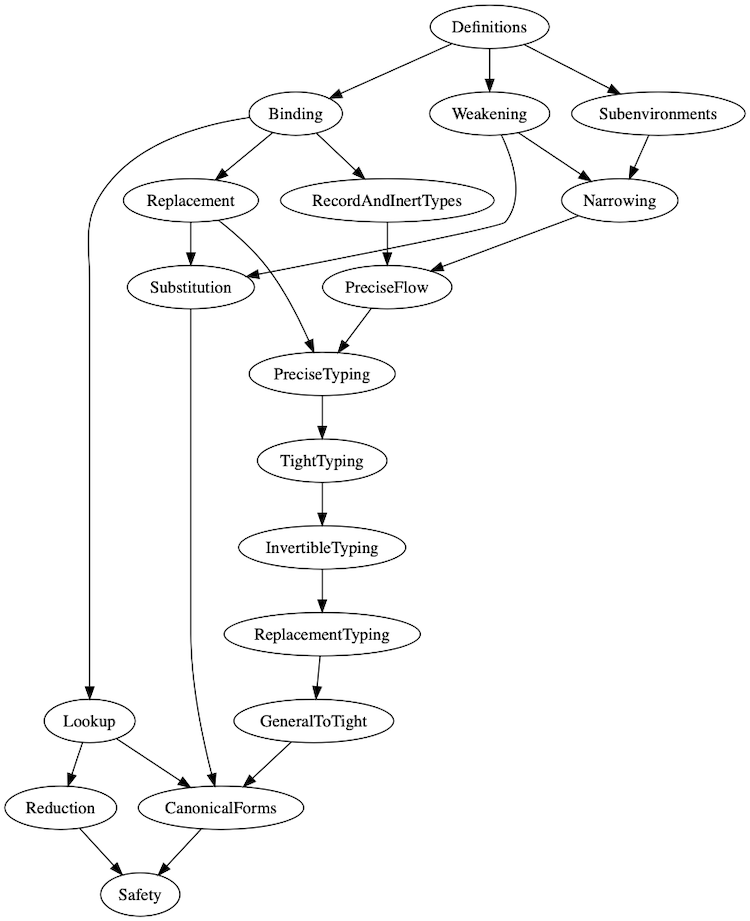}
    \caption{Dependency between Coq proof files}\label{fig:coqdepgraph}
\end{figure}

Figure~\ref{fig:coqdepgraph} shows a dependency graph between the Coq modules.

\end{document}